\documentclass[sigconf]{acmart}
\usepackage{caption}
\usepackage{subcaption}
\usepackage[]{algorithm2e,algorithmic}
\usepackage{amsmath}
\usepackage{flushend}
\makeatletter
\newcommand{\removelatexerror}{\let\@latex@error\@gobble}
\makeatother
\newcommand{\myparagraph}[1]{\paragraph{{\bf #1}}}

\newcommand\addauthornote[1]{%
  \if@ACM@anonymous\else
    \g@addto@macro\addresses{\@addauthornotemark{#1}}%
  \fi}

\AtBeginDocument{%
  \providecommand\BibTeX{{%
    \normalfont B\kern-0.5em{\scshape i\kern-0.25em b}\kern-0.8em\TeX}}}

\setcopyright{acmcopyright}
\copyrightyear{2021}
\acmYear{2021}
\acmConference[]{}{}{}
\acmBooktitle{}
\begin{document}

\title{Deep Retrieval: Learning A Retrievable Structure for Large-Scale Recommendations}

\author{Weihao Gao$^\dagger$, Xiangjun Fan$^\dagger$, Chong Wang$^\dagger$, Jiankai Sun, Kai Jia, Wenzhi Xiao\\ Ruofan Ding, Xingyan Bin, Hui Yang, Xiaobing Liu}\thanks{Corresponding authors: \{weihao.gao, xiangjun.fan, chong.wang\}@bytedance.com. \\ $\dagger$ indicates equal contributions.}
\affiliation{%
\institution{ByteDance Inc.}
\country{}
}

\renewcommand{\shortauthors}{Gao and Fan, et al.}

\begin{abstract}
One of the core problems in large-scale recommendations is to retrieve top relevant candidates accurately and efficiently, preferably in sub-linear time. Previous approaches are mostly based on a two-step procedure: first learn an inner-product model, and then use some approximate nearest neighbor (ANN) search algorithm to find top candidates. In this paper, we present Deep Retrieval (DR), to learn a retrievable structure directly with user-item interaction data (e.g. clicks) without resorting to the Euclidean space assumption in ANN algorithms. DR's structure encodes all candidate items into a discrete latent space. Those latent codes for the candidates are model parameters and learnt together with other neural network parameters to maximize the same objective function. With the model learnt, a beam search over the structure is performed to retrieve the top candidates for reranking. Empirically, we first demonstrate that DR, with sub-linear computational complexity, can achieve almost the same accuracy as the brute-force baseline on two public datasets. Moreover, we show that, in a live production recommendation system, a deployed DR approach significantly outperforms a well-tuned ANN baseline in terms of engagement metrics. To the best of our knowledge, DR is among the first non-ANN algorithms successfully deployed at the scale of hundreds of millions of items for industrial recommendation systems. 
\end{abstract}

\begin{CCSXML}
<ccs2012>
 <concept>
  <concept_id>10010520.10010553.10010562</concept_id>
  <concept_desc>Computer systems organization~Embedded systems</concept_desc>
  <concept_significance>500</concept_significance>
 </concept>
 <concept>
  <concept_id>10010520.10010575.10010755</concept_id>
  <concept_desc>Computer systems organization~Redundancy</concept_desc>
  <concept_significance>300</concept_significance>
 </concept>
 <concept>
  <concept_id>10010520.10010553.10010554</concept_id>
  <concept_desc>Computer systems organization~Robotics</concept_desc>
  <concept_significance>100</concept_significance>
 </concept>
 <concept>
  <concept_id>10003033.10003083.10003095</concept_id>
  <concept_desc>Networks~Network reliability</concept_desc>
  <concept_significance>100</concept_significance>
 </concept>
</ccs2012>
\end{CCSXML}

\ccsdesc[300]{Computing methodologies~Machine Learning}
\ccsdesc[300]{Mathematics and computing~Probability and statistics}

\keywords{deep retrieval, recommendation systems}


\maketitle

\section{Introduction}
\label{sec:intro}
Recommendation systems have gained great success in various commercial applications for decades. The objective of these systems is to return relevant items from a corpus based on user features and historical behaviors. One of the early successful techniques is the collaborative filtering (CF), based on the simple idea that similar users may prefer similar items. Item-based collaborative filtering (Item-CF)~\citep{sarwar2001item} extends this idea by considering the similarities between items and items, which lays the foundation for Amazon's recommendation system~\citep{linden2003amazon}. 

In the Internet era, the size of users and items on popular content platforms rapidly grow to tens to hundreds of millions. 
%making it more challenging to design accurate recommendation systems. 
The scalability, efficiency as well as accuracy, are all challenging problems in the design of modern recommendation systems.
%One of the early successful techniques of recommendation systems is the collaborative filtering (CF), which makes predictions based on the simple idea that similar users may prefer similar items. Item-based collaborative filtering (Item-CF)~\citep{sarwar2001item} extends the idea by considering the similarities between items and items, and later lays a foundation for Amazon's recommendation system~\citep{linden2003amazon}. 
%However, collaborative filtering algorithms could not solve cold-start problem or sparsity of data well, which are key challenges in modern recommendation systems.
Recently, vector-based retrieval methods have been widely adopted. The main idea is to embed users and items in a latent vector space, and use the inner product of vectors to represent the preference between users and items. Representative vector embedding methods include matrix factorization (MF)~\citep{mnih2008probabilistic,koren2009matrix}, factorization machines (FM)~\citep{rendle2010factorization}, DeepFM~\citep{guo2017deepfm}, Field-aware FM (FFM)~\citep{juan2016field}, etc. However, when the number of items is large, the cost of brute-force computation of the inner products for all items can be prohibitive. Thus, approximate nearest neighbors (ANN) or maximum inner product search (MIPS) algorithms are usually used to retrieve top relevant items when the corpus is too large. 
Efficient MIPS or ANN algorithms include tree-based algorithms~\citep{muja2014scalable,houle2014rank}, locality sensitive hashing (LSH)~\citep{shrivastava2014asymmetric,spring2017new}, product quantization (PQ)~\citep{jegou2010product,ge2013optimized}, hierarchical navigable small world graphs (HNSW)~\citep{malkov2018efficient}, etc. 

Despite their success in real world applications, vector-based methods with ANN or MIPS have two main disadvantages: (1) the inner product structure of user and item embeddings might not be sufficient to capture the complicated structure of user-item interactions~\citep{he2017neural}. (2) ANN or MIPS is designed to approximate the learnt inner product model, not directly optimized for the user-item interaction data. As a solution, tree-based models~\citep{ zhu2018learning,zhu2019joint,zhuo2020learning} have been proposed to address these issues. One potential issue is of these methods is that each item is mapped to a leaf node in the tree, 
%Although the learning objectives for model parameters and the tree structure are well-aligned, resulting in improving the accuracy, the number of parameters in these models are proportional to the number of clusters
making the tree structure itself difficult to learn. Data available at the leaf level can be scarce and might not provide enough signal to learn a good tree structure at a finer level for a recommendation system with hundreds of millions of items. An efficient and easy-to-learn retrieval system is still very much needed for large-scale recommendations. %However, the structure of the tree is discrete, hence can not be trained as efficient as the continuous model parameters. Therefore, these algorithm are not end-to-end trainable and are difficult to be fully deployed on modern deep learning platforms.

\begin{figure*}[htpb]
  \centering
  \subcaptionbox{\label{fig:structure model}}{%
    \includegraphics[width=0.4\textwidth]{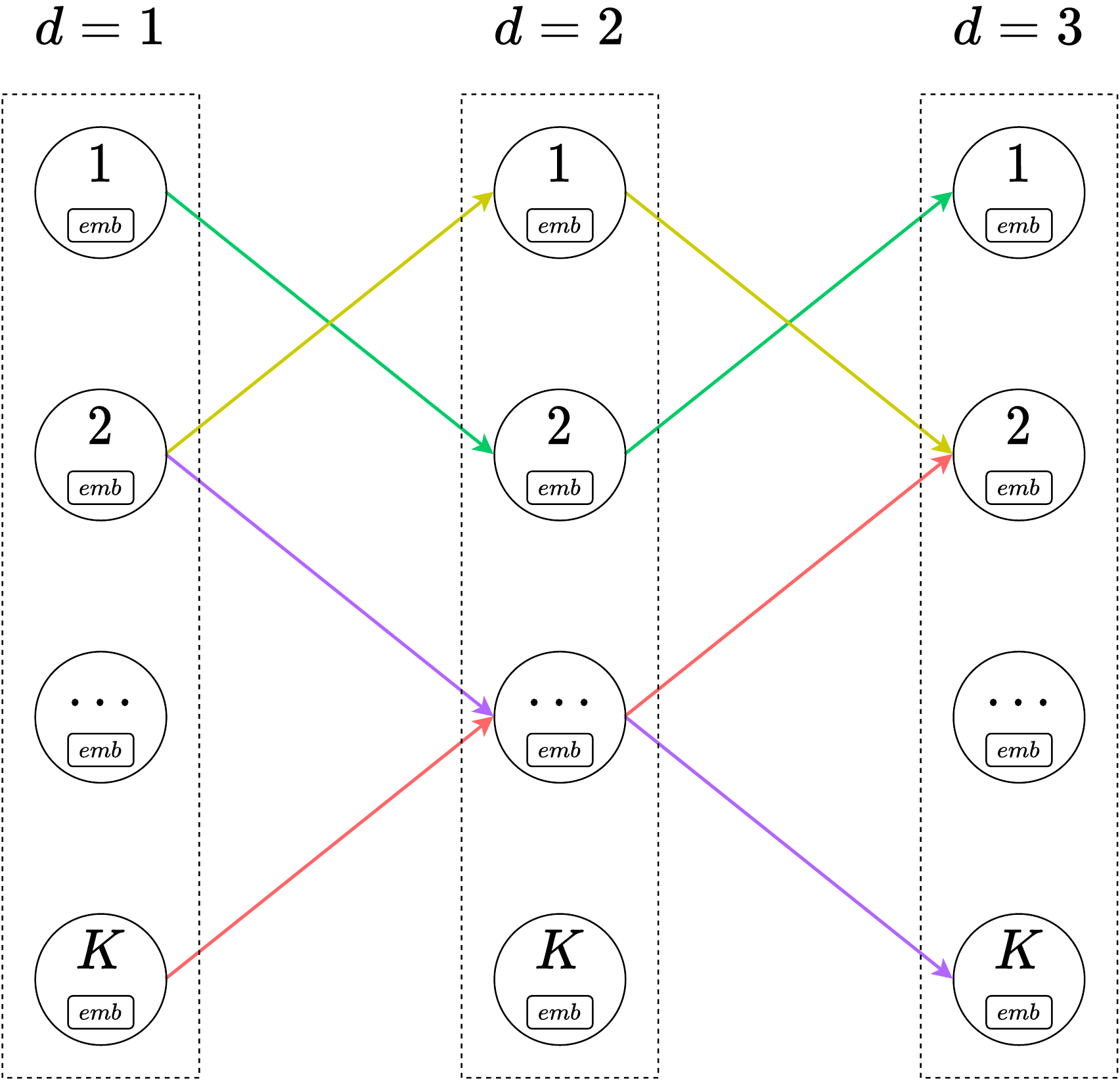}
  }
  \hspace{0.4in}
  \subcaptionbox{\label{fig:structure_los}}{%
    \includegraphics[width=0.5\textwidth]{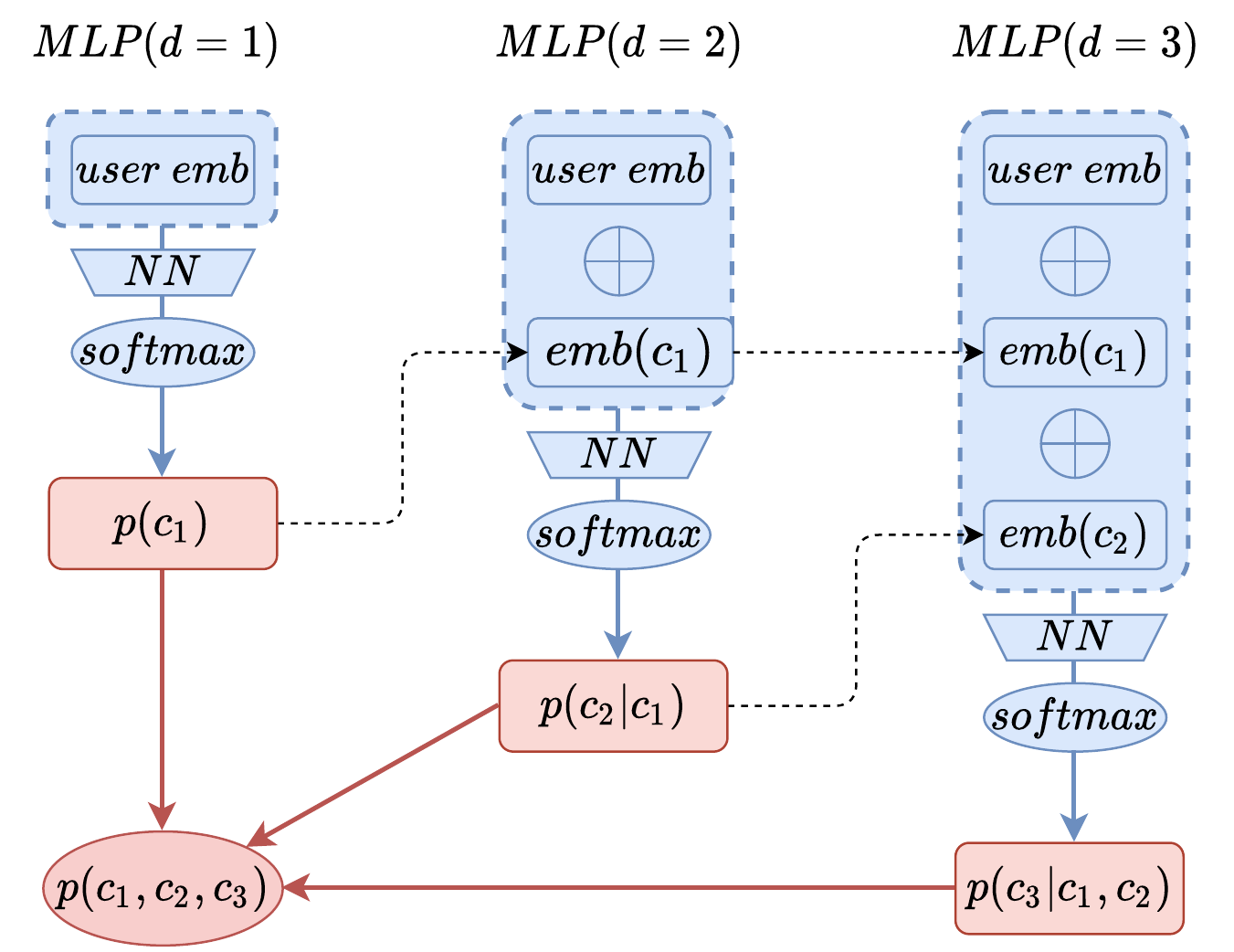}
  }
  \caption
        {(a) Consider a structure with width $K=100$ and depth $D=3$, and an item encoded by a path of $[36,27,20]$. This path denotes that the item is assigned to the $(1,36)$, $(2, 27)$, $(3,20)$ indices of the $K \times D$ matrix. In the figure, arrows with the same color form a path. Different paths could intersect with each other by sharing a common index in a layer. (b) Flow diagram of the process for constructing $p(c|x, \theta)$, the probability of path $c=[c_1, c_2, c_3]$ given input $x$.} 
        \label{fig:ColdStart_SequentialModels}
\end{figure*}

In this paper, we proposed Deep Retrieval (DR) to learn a retrievable structure in an end-to-end manner from data. DR uses a $K \times D$ matrix as shown in Fig.~\ref{fig:structure model} for indexing, motivated by the ideas in~\citet{chen2018learning} and~\citet{van2017neural}, and we have introduced interdependence among the code in different $d \in \{1,...,D\}$. In this structure, we define a path $c$ as the forward index traverses over matrix columns. Each path is of length $D$ with index value range $\{1, 2, \dots, K\}$. There are $K^D$ possible paths and each path can be interpreted as a cluster of items. We learn a probability distribution over the paths given the user inputs, jointly with a mapping from the items to the paths. In the serving stage, we use beam search to retrieve the most probable paths and the items associated with those paths. There are two major characteristics in designing the DR structure. 
\begin{enumerate}
    %\item There is no ``leaf node'' as that in a tree, so the data scarcity problem of learning the tree based model can be largely avoided in DR's structure.
    \item DR serves the purpose for ``retrieval'', not ranking. Since there is  no ``leaf node'' in DR's structure, items within a path are indistinguishable for retrieval purpose, mitigating the data scarcity problem for ranking these items. DR can be easily scaled up to hundreds of millions items. 
    \item Each item is designed to be indexed by more than one path, meaning that two items can share some paths but differ in other paths. This design can be naturally realized using our probabilistic formulation of the DR model as we will show later. This multiple-to-multiple encoding scheme between items and paths differs significantly with the one-to-one mapping used in ealier designs of tree-based structures. 
\end{enumerate}
In training, the item paths are also model parameters and are learned together with other neural network parameters of the structure model using an expectation-maximization (EM) type algorithm~\citep{dempster1977maximum}. %The entire training process is end-to-end and can be easily deployed for large-scale content platforms.

%In terms of model capability, the structure model enables DR to have much more clusters than its parameters, which solves the problem of scarce data in TDM and JTM. In addition, the multiple-to-multiple encoding scheme enables DR to learn more complicated relationships among users and items. Experiments show the benefit of being able to learn such complicated relationships.

%For example, an item related to chocolate may be encoded by $[36,204,105]$ and another item related to cake may be encoded by $[36,227,120]$.

% \begin{figure}
%     \centering
%     \includegraphics[width=0.55\textwidth]{figures/DR_Indexing_Illustration.pdf}
%     \caption{An illustration of DR learned item indexing. Consider a structure with width $K=100$ and depth $D=3$. Assuming an item is encoded by length-$D$ vector $[36,27,20]$, which is called a ``code'' or a ``path''. The path denotes that the item is assigned to the $(1,36)$, $(2, 27)$, $(3,20)$ indices of the $K \times D$ matrix. In the figure, arrows with the same color form a path. Different paths could intersect with each other by sharing the same index at some layer.}
%     \label{fig:structure model}
% \end{figure}

%The advantage of DR is multi-folded.  

% \begin{itemize*}
%     \item 
%     \item 
%     %We also develop an online EM algorithm for the online learning scenario where only streaming training data are available.
%     %\item In Section~\ref{sec:architecture}, we introduce the training and online serving architectures for Deep Retrieval. The training architecture is not only designed for DR, but also suitable to any other models which combines gradient-based training and EM algorithm.

% \end{itemize*}

% \begin{itemize}
%     \item Recommendation is a core problem in machine learning... (references)
%     \item Inner-product based algorithms... (references, limitations)
%     \item Tree based methods (TDM, JTM, limitations)
%     \item Motivated by this, we propose to encode each item using $K \times D$ matrix. (Figure 1). 
%     \item The rest of the paper is organized as follows...
% \end{itemize}

\myparagraph{Related Works}

Here we briefly review some related works and discuss their connections to DR. Among numerous large-scale recommendation algorithms, the closest works to ours are the tree-based methods, including tree-based deep model (TDM)~\cite{zhu2018learning}, JTM~\cite{zhu2019joint} , OTM~\cite{zhuo2020learning} and AttentionXML~\cite{you2019attentionxml}. Tree-based methods map each item to a leaf node in a tree-based structure model and learn an objective function for the tree structure and the model parameters jointly. As discussed above, learning a good tree structure for hundreds of millions of items can be difficult.

Another line of research attempts to encode the candidates in a discrete latent space. Vector Quantised-Variational AutoEncoder(VQ-VAE)~\cite{van2017neural} uses a similar $K \times D$ embedding space to encode items, but focuses more on encoding data such as image or audio, whereas DR focuses on clustering items based on user-item interactions. HashRec~\cite{kang2019candidate} and Merged-Averaged Classifiers via Hashing~\cite{medini2019extreme} use multi-index hash functions to encode candidate items in large recommendation systems. Compared to HashRec and MACH, DR uses a more complicated structure model with dependencies among layers and a different beam search based inference algorithm.

More complicated structure models have been used in extreme classification. LTLS~\cite{jasinska2016log} and W-LTLS~\cite{evron2018efficient} utilize directed acyclic graphs (DAG) as structure models. Probabilistic classifier chains~\cite{dembczynski2010bayes} are also used as structure models for multi-label classification. We note that the number of labels in these extreme classification applications is at most tens of thousands, whereas DR can be applied to industrial recommendation systems containing hundreds of millions of items.

\myparagraph{Organization}

The rest of the paper is organized as follows. In Section~\ref{sec:model}, we describe the structure model and its training objective function in detail. We then introduce a beam search algorithm to find candidate paths during the inference stage. In Section~\ref{sec:learning}, we introduce the EM algorithm for training neural network parameters and paths of items jointly. In Section~\ref{sec:exp}, we first demonstrate the performance of DR on two public datasets: MovieLens-20M\footnote{\url{https://grouplens.org/datasets/movielens}} and Amazon books\footnote{\url{http://jmcauley.ucsd.edu/data/amazon}}. Experiment results show that DR can achieve almost the brute-force accuracy with sub-linear computational complexity. Moreover, in Section~\ref{sec:live-exp}, we test the performance of DR on a live production recommendation system with hundreds of millions of users and items and show that it significantly outperforms a well-tuned ANN baseline in terms of engagement metrics in A/B test. In Section~\ref{sec:discussion}, we conclude the paper and discuss several possible future research directions.

\section{Deep retrieval: Learning A Retrievable Structure from data}
\label{sec:model}
In this section, we introduce DR in detail. First, we establish its basic probability formulation. We then extend it to a multi-path mechanism that enables DR to capture multi-aspect properties of items. We then introduce a penalization design that prevents collapsing in allocating items to different paths. Next, we describe a beam search algorithm for retrieval. Finally, we present the multi-task joint training procedure of DR with a reranking model. 

\subsection{The DR Model}
\myparagraph{The basic model.}
The basic DR model consists of $D$ layers, each with $K$ nodes. In each layer, we use a multi-layer perceptron\footnote{Other complex neural network architectures such as recurrent neural networks  can also be applied here. For simplicity, we only use MLP in our experiments.} (MLP) and $K$-class softmax to output a distribution over its $K$ nodes. 
%Each layer takes an input vector and outputs a probability distribution over $\{1, 2, \dots, K\}$ based on parameters $\theta$.
Let $\mathcal{V} = \{1, \dots, V\}$ be the labels of all items and an item-to-path mapping $\pi: \mathcal{V} \to [K]^{D}$. Here a path is the forward index traverse over matrix columns as shown in Fig.~\ref{fig:structure model}. For simplicity, we assume that the mapping $\pi$ is given in this section (We will introduce the algorithm for learning the mapping $\pi$ later in Section~\ref{sec:learning}.) In addition, we assume an item can only be mapped to one path for now and leave the multi-path setting to the next section.

Given a pair of training sample $(x,y)$, which denotes a positive interaction (click, convert, like, etc.) between a user $x$ and an item $y$,  as well as the path $c = (c_1, c_2, \dots, c_D)$ associated with the item $y$, i.e. $\pi(y)=c$, the path probability $p(c|x, \theta)$ is constructed layer by layer as follows (see Fig.~\ref{fig:structure_los} for a flow chart),

% \begin{figure}
%     \centering
%     \includegraphics[width=0.9\textwidth]{figures/DR_Structure_Illustration.pdf}
%     \caption{Flow chart showing the process for constructing the probability $p(c|x, \theta)$.}
%     \label{fig:structure_loss}
% \end{figure}

\begin{itemize}
    \item The first layer takes the user embedding ${\rm emb}(x)$ as input, and outputs a probability $p(c_1 | x, \theta_1)$ over the $K$ nodes of the first layer, based on parameters $\theta_1$ .
    \item From the second layer onward, we concatenate the user embedding ${\rm emb}(x)$ and the embeddings of all the previous layers ${\rm emb}(c_{d-1})$ (called path embeddings) as the input of MLP, which outputs $p(c_d| x, c_1, \dots, c_{d-1}, \theta_d)$ over the $K$ nodes of layer $d$, based on parameters $\theta_d$.
    \item The probability of path $c$ given user $x$ is the product of the probabilities of all the layers' outputs: 
    \begin{align}
        p(c | x, \theta) = \prod_{d=1}^D p(c_d | x, c_1, \dots, c_{d-1}, \theta_d). \label{eq:item_to_path}
    \end{align}
\end{itemize}
Given a set of $N$ training samples $\{(x_i, y_i)\}_{i=1}^N$, the log likelihood function of the structure model is
\begin{align*}
    \mathcal{Q}_{\rm str}(\theta, \pi) =  \sum_{i=1}^N \log p(\pi(y_i) | x_i, \theta) .%= \sum_{i=1}^N \sum_{d=1}^D \log p(\pi(y_i)_d | x_i, \pi(y_i)_1 , \dots, \pi(y_i)_{d-1} , \theta_d).
    %\label{eq:structure_loss}
\end{align*}
The size of the input vector of layer $d$ is the embedding size times $d$, and the size of the output vector is $K$. The parameters of layer $d$ have a size of $\Theta(Kd)$. The parameters $\theta$ contain the parameters $\theta_d$'s in all layers, as well as the path embeddings. The number of parameters in the entire model has an order of $\Theta(KD^2)$, which is significantly smaller than the number of possible paths $K^D$ when $D \geq 2$. 

\myparagraph{Multi-path extension.}
In tree-based deep models~\citep{zhu2018learning,zhu2019joint} as well as the structure model we introduced above, each item belongs to only one cluster/path, limiting the capacity of the model for expressing multi-aspect information in real data. For example, an item related to kebab could belong to a ``food'' cluster. An item related to flowers could belong to a ``gift'' cluster. However, an item related to chocolate or cakes could belong to both clusters in order to be recommended to users interested in either food or gifts. In real world recommendation systems, a cluster might not have an explicit meaning such as food or gifts, but this example motivates us to assign each item to multiple clusters. In DR, we allow each item $y_i$ to be assigned to $J$ different paths $\{c_{i, 1}, \dots, c_{i, J}\}$. Let $\pi: \mathcal{V} \to [K]^{D \times J}$ be the mapping from items to multiple paths. The multi-path structure objective is straightforwardly defined as
\begin{align*}
    \mathcal{Q}_{\rm str}(\theta, \pi) = \sum_{i=1}^N \log \left( \sum_{j=1}^J p(c_{i,j} = \pi_j(y_i)| x_i, \theta) \right),
   % \label{eq:multi_path_structure_loss}
\end{align*}
where the probability belonging to multiple paths is the summation of the probabilities belonging to individual paths.

% \begin{algorithm}
%     \caption{Beam search algorithm}
%     \label{algo:beam_search}
%     \begin{algorithmic}
%         \STATE {\bf Input} user $x$, structure model with parameter $\theta$, beam size $B$.
%         \STATE Let $C_1 = \{c_{1,1}, \dots, c_{1,B}\}$ be top $B$ entries of $\{p(c_1 | x, \theta): c_1 \in \{1, \dots, K\}\}$.
%         \FOR{$d = 2$ to $D$}
%             \STATE Let $C_d = \{(c_{1,1}, \dots, c_{d,1}), \dots, (c_{1,B}, \dots, c_{d,B})\}$ be the top $B$ entries of the set of all successors of $C_{d-1}$ defined as follows.
%             \begin{equation*}
%                 \{p(c_1, \dots, c_{d-1}|x, \theta) p(c_d | x, c_1, \dots, c_{d-1}, \theta): (c_1, \dots, c_{d-1}) \in C_d, c_d \in \{1, \dots, K\}\}.
%             \end{equation*}
%         \ENDFOR
%         \STATE {\bf Output} $C_D$, a set of $B$ paths.
%     \end{algorithmic}
% \end{algorithm}

% \begin{itemize}
%     \item Describe structure model and structure loss in detail (Figure 2, more detailed illustration of structure model than Figure 1)
%     \item Describe beam search algorithm for finding candidate paths
%     \item Multi-path. Use chocolate, flower and sandwich example.
% \end{itemize}

\myparagraph{Penalization on size of paths.}
%Overfitting is likely to happen if we do not apply any penalization for Algorithm~\ref{alg:structure_learning}. Imagine that the structure model is overfitted and gives a particular path $(0,0,0)$ a very high probability, for	 any input. Then in M-step, all the items will be assigned to path $(0,0,0)$, making the model fail to cluster the items. In order to prevent overfitting, we introduce a penalization term on the size of the paths.
Directly optimizing $\mathcal{Q}_{\rm str}(\theta, \pi)$  w.r.t. the item-to-path mapping $\pi$ could fail to allocate different items into different paths (we did observe this in practice). In an extreme case, we could allocate all items into a single path and the probability of seeing this single path given any user $x$  is 1 as in Eq.~\ref{eq:item_to_path}. This is because there is only one available path to choose from. However, this will not help on retrieving the top candidates since there is no differentiation among the items. We must regulate the possible distribution of $\pi$ to ensure diversity. We introduce the following penalized likelihood function,
\begin{align*}
    \mathcal{Q}_{\rm pen}(\theta, \pi) = \mathcal{Q}_{\rm str}(\theta, \pi) - \alpha \cdot \sum_{c \in [K]^D} f(|c|),  %\,\notag\\
    %&=\sum_{v=1}^V \left( N_v \log \left( \sum_{j=1}^J s[v, \pi_j(v)]\right) - \log N_v\right) - \alpha \cdot \sum_{c \in [K]^D} f(|c|).
    %\label{eq:penalized_loss}
\end{align*}
where $\alpha$ is the penalty factor, $|c|$ denotes the number of items allocated in path $c$ and $f$ is an increasing and convex function. A quadratic function $f(|c|) = |c|^2/2$ controls the average size of paths, and higher order polynomials penalize more on larger paths. In our experiments, we use $f(|c|) = |c|^4/4$. 
%It's worth mentioning that the penalty is only applied in M-step, not during training of continuous parameters $\theta$.

\subsection{Beam Search for Inference} In the inference stage, we want to retrieve items from the DR model, given user embeddings as input. To this end, we use the beam search algorithm~\citep{reddy1977speech} to retrieve most probable paths. In each layer, the algorithm selects top $B$ nodes from all the successors of the selected nodes from the previous layer. Finally it returns $B$ top paths in the final layer. When $B=1$, this becomes the greedy search. In each layer, choosing top $B$ from $K \times B$ candidates has a time complexity of $O(KB \log B)$. The total complexity is $O(DKB \log B)$, which is sub-linear with respect to the total number of items $V$. We summarize the procedure in Algorithm~\ref{algo:beam_search}.

\begin{algorithm}[th]
        {\bf Input}: user $x$, model parameter $\theta$, beam size $B$. \\
        Let $C_1 = \{c_{1,1}, \dots, c_{1,B}\}$ be top $B$ entries of $\{p(c_1 | x, \theta): c_1 \in \{1, \dots, K\}\}$. \\
        \For{$d = 2$ to $D$} {
            Let $C_d = \{(c_{1,1}, \dots, c_{d,1}), \dots, (c_{1,B}, \dots, c_{d,B})\}$ be the top $B$ entries of the set of all successors of $C_{d-1}$ defined as follows. \\
                $p(c_1, \dots, c_{d-1}|x, \theta) p(c_d | x, c_1, \dots, c_{d-1}, \theta)$, where $(c_1, \dots, c_{d-1}) \in C_d, c_d \in \{1, \dots, K\}$.
        }
       {\bf Output}: $C_D$, a set of $B$ paths.
      \vspace{0.1in}
    \caption{Beam search algorithm}
    \label{algo:beam_search}
\end{algorithm}

\subsection{Multi-task Learning and Reranking with Softmax Models}
%In the structure model described above, we only make use of positive instances and maximize the log probability of positive instances. This makes the model able to recognize ``good'' items, but unable to distinguish ``good'' items from ``bad'' ones. So the model is good at retrieval but not ranking. To this end, we introduced a rerank model $\mathcal{Q}_{\rm softmax}$ trained with sampled softmax objective described as 
The number of items returned by DR beam search is much smaller than the total number of items, but often not small enough to serve the user request,
%\footnote{How to make DR more effective in returning a smaller number of items is left as future research.} 
so we need to rank those retrieved items.
However, since each path in DR can contain more than one item, it is not easy to differentiate these items using DR alone. We choose to tackle this by jointly training a DR model with a reranker:
%In the inference stage, we introduce a beam search algorithm to select candidate paths based on user input. 
\begin{align*}
    \mathcal{Q}_{\rm softmax} = \sum_{i=1}^N \log p_{\rm softmax}(y=y_i|x_i),%\frac{\exp(x_i^\top y_i)}{\sum_{j=1}^M \exp(f_{\phi}(x'_{i,j}, y'_{i,j}))}
\end{align*}
where $p(y=y_i|x_i)$ is a softmax model with output size $V$. This is trained with the sampled softmax algorithm.
%By sharing the input with the softmax model, we teach DR to distinguish positive instances from random negative instances. 
The final objective is given by $\mathcal{Q} = \mathcal{Q}_{\rm pen} + \mathcal{Q}_{\rm softmax}$. After performing a beam search to retrieval a set of candidate items, we rerank those candidates to obtain the final top candidates. Here we use a simple softmax model, but we can certainly replace it by a more complex one for better ranking performance.

% Based on the experiments we conducted in this paper, we found that jointly training DR with a softmax classification model greatly improves the performance. We conjecture that this is because the paths for the items are randomly assigned in the beginning, leading to increased difficulty for optimization. By sharing the inputs with an easy-to-train softmax model, we are able to give the structure model an uplift in the optimization direction. So the final objective we are maximizing is
% \begin{align*}
% \mathcal{Q} = \mathcal{Q}_{\rm str} + \mathcal{Q}_{\rm softmax}.
% \end{align*}
% 
    
\section{Learning with the EM algorithm}
\label{sec:learning}
In the previous section, we introduced the structure model in DR and its objective to be optimized. The objective is continuous with respect to the neural network parameters $\theta$, which can be optimized by any gradient-based optimizer. However, the objective involving the item-to-path mapping $\pi$ is discrete and can not be optimized by a gradient-based optimizer. As this mapping acts as the ``latent clustering'' of items, this motivates us to use an EM-style algorithm to optimize the mapping and other parameters jointly. %In this section, we describe the EM algorithm in detail and introduce a penalization term to prevent overfitting. 
%Moreover, an online variant of the EM algorithm is introduced for online training scenario.
%\subsection{EM Algorithm for Joint Training}

In general, the EM-style algorithm for DR is summarized as follows. We start with initializing mapping $\pi$ and other parameters randomly. Then at the $t^{\rm th}$ epoch,
\begin{enumerate}
    \item E-step: for a fixed mapping $\pi^{(t-1)}$, optimize parameter $\theta$ using a gradient-based optimizer to maximize the structure objective $\mathcal{Q}_{\rm pen}(\theta, \pi^{(t-1)})$.
    \item M-step: update mapping $\pi^{(t)}$ to maximize the same structure objective $\mathcal{Q}_{\rm pen}(\theta, \pi)$.
\end{enumerate}
Since the E-step is similar to any standard stochastic optimization, we will focus on the M-step here. For simplicity, we first consider the objective $\mathcal{Q}_{\rm str}$ without the penalization. Given a user-item training pair $(x_i, y_i)$, let the set of paths associated with the item be $\{\pi_1(y_i), \dots, \pi_J(y_i)\}$. For a fixed $\theta$, we can rewrite $\mathcal{Q}_{\rm str}(\theta, \pi)$ as
\begin{align}
    \mathcal{Q}_{\rm str}(\theta, \pi) &=  \sum_{i=1}^N \log \left( \sum_{j=1}^J p(\pi_j(y_i) | x_i, \theta) \right) \,\notag \\
    &=  \sum_{v=1}^V \left( \sum_{i: y_i=v} \log \left( \sum_{j=1}^J p(\pi_j(v) | x_i, \theta) \right) \right),
   \label{eq:score}
\end{align}
where the outer summation is over all items $v \in \mathcal{V}$ and the inner summation is over all appearances of item $v$ in the training set. We now consider maximizing the objective function over all possible mappings $\pi$. However, for an item $v$, there are $K^D$ number of possible paths so we could not enumerate it over all $\pi_j(v)$'s. %Instead, for each appearance, we only record the probability scores for the top paths using beam search and leave the rest with zero scores. 
%Unfortunately, the top paths for item $v$ in different appearances may not be the same due to the difference of user and context embeddings, which renders an unstable function due to the presence of $\log(p)+\log(0)$ in Equation~\eqref{eq:score}\footnote{We also tried to use a small value. But it is difficult to figure out what the small value should be.}. 
%Still the current form is still far from tractable for updating the mapping $\pi$.
We make the follow approximation to further simplify the problem since Eq.~\ref{eq:score} is not easily solvable. We use an upper bound  $\sum_{i=1}^N \log p_i \leq N (\log {\sum_{i=1}^N p_i } - \log N)$ to obtain \footnote{Theoretically, maximizing an upper bound approximation to the true objective, as we do here, does not provide guarantees to the quality of the solution. This is different from the common practice of maximizing a lower bound. Nevertheless, we observed decent performance of our proposed method in practice.}
\begin{align*}
    &\mathcal{Q}_{\rm str}(\theta, \pi) \leq \overline{\mathcal{Q}}_{\rm str}(\theta, \pi) \,\notag\\
    &= \sum_{v=1}^V \left( N_v \log \left( \sum_{j=1}^J \sum_{i: y_i=v}  p(\pi_j(v) | x_i, \theta) \right) - \log N_v\right),
\end{align*}
where $N_v = \sum_{i=1}^N \mathbb{I} [i:y_i=v]$ denotes the number of occurrences of $v$ in the training set which is independent of the mapping. We define the following score function,
\begin{align*}
s[v, c] \triangleq \sum_{i: y_i=v}  p(c | x_i, \theta).
\end{align*}
Intuitively, $s[v, c]$ can be understood as the accumulated importance score of allocating item $v$ to path $c$. In practice, it is impossible to retain all scores as the possible number of paths $c$ is exponentially large, so we only retain a subset of $S$ paths with largest scores through beam search and set the rest of the scores as $0$.

\myparagraph{Remark on estimating $s[v, c]$ with streaming training.}  Given the huge amount of continuously arriving data in real production recommendation systems, we typically have to use streaming training---processing the data only once ordered by their timestamps.
So we choose to estimate scores $s[v,c]$ in a streaming fashion inspired by the approach in~\citet{metwally2005efficient}. The basic idea is to keep tracking a list of top $S$ important paths. For item $v$, suppose we have recorded score list $s[v, c_{1:S}]$. After we obtain a new list of scores,  $s'[v, c'_{1:S}]$, from a new training instance containing $v$, we update the score list $s[v, c_{1:S}]$ as follows,
\begin{enumerate}
    \item Let ${\rm min}\_{\rm score} = \min_i s[v, c_i] $.
    \item Create a union set set $A=c_{1:S}\bigcup c'_{1:S}$. 
    \item For each $c \in A$:
    \begin{enumerate}
        \item If $c\in c_{1:S}$ and $c\in c'_{1:S}$, set $s[v, c] \leftarrow \eta\ s[v, c] + s'[v, c]$.
         \item If $c\notin c_{1:S}$ and $c\in c'_{1:S}$, set $s[v, c] \leftarrow \eta\ {\rm min}\_{\rm score}  + s'[v, c]$.
        \item If $c\in c_{1:S}$ and $c\notin c'_{1:S}$, set $s[v, c] \leftarrow \eta\ s[v, c]$.
    \end{enumerate}
    \item Choose $S$ largest values of $s[v, c]$ from $c\in A$ to form the new score vector $s[v, c_{1:S}]$.
\end{enumerate}
\begin{figure}
    \centering
    %\vspace{-0.2in}
    \includegraphics[width=0.45\textwidth]{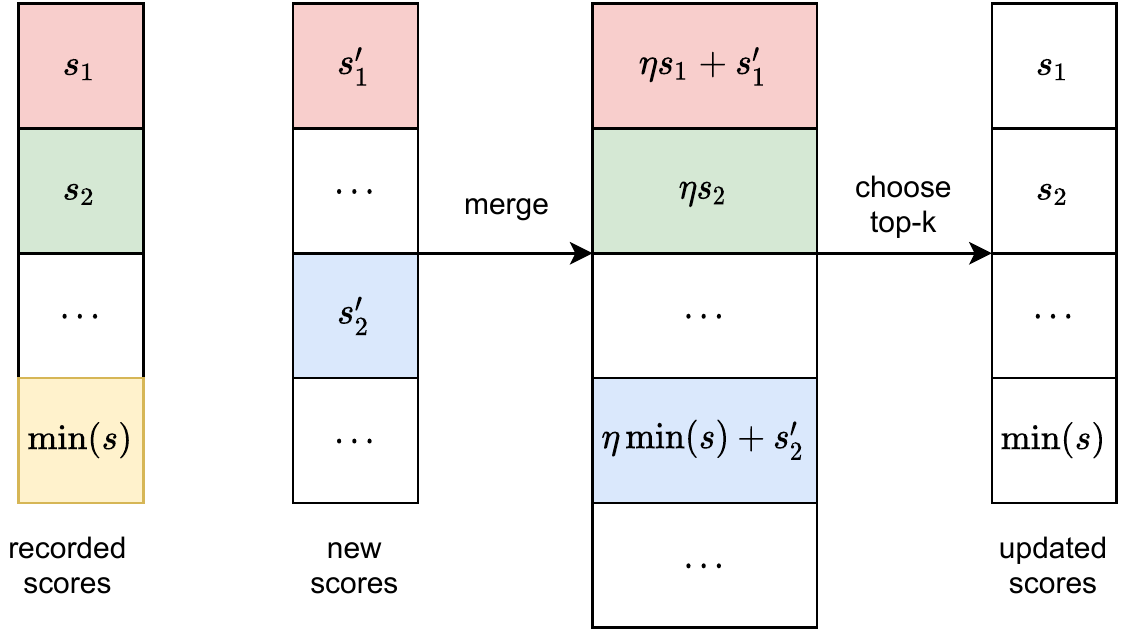} 
    %\vspace{-0.3in}
    \caption{Illustration of the score updating process. $min(s)$ is used to increase the exploration of new paths in training. Blocks with the same colors indicate the same paths.}
    \label{fig:merge_score}
    \vspace{-0.2in}
\end{figure}
Here $\eta$ is a decay factor to account for the streaming estimation and we use $\eta=0.999$ in our experiments. As shown in Fig.~\ref{fig:merge_score}, if the path (in red) appears both in the recorded list and the new list, we update the score by a discounted rolling sum. If the path (in green) appears only in the recorded list but not in the new list, we simply discount the score. If the path (in blue) appears only in the new list but not in the recorded list, we use ${\rm min}\_{\rm score}$ instead of $0$ as its score in the recorded list. This method increases the likelihood of exploring new paths in the streaming fashion, and we found it is important for the approach to work well.

\paragraph{Solving the M-step using coordinate descent.} Given score $s[v, c]$ estimated, now we consider the objective $\mathcal{Q}_{\rm pen}$ with penalization. This leads to the following surrogate function in the M-Step,
\begin{align*}
    \arg\max_{\{\pi_{j}(v)\}_{j=1}^J} &&\sum_{v=1}^V \left( N_v \log \left( \sum_{j=1}^J s[v, \pi_j(v)]\right) - \log N_v\right) \,\notag\\
    &&- \alpha \cdot \sum_{c \in [K]^D} f(|c|).
\end{align*}
Notice that there is no closed-form solution for this maximization problem. Hence, we utilize the coordinate descent algorithm, 
%We illustrate the coordinate descent algorithm used in path assignment with penalty in detail. Recall that in penalized M-step, we want to maximize the following objective over all assignments $\{\pi_j(v)\}_{j=1}^J$'s.
%
%\begin{equation}
%    \argmax_{\{\pi_{j}(v)\}_{j=1}^J} \sum_{v=1}^V \left( N_v \log \left( \sum_{j=1}^J s[v, \pi_j(v)]\right) - \log N_v\right) - \alpha \cdot \sum_{c \in [K]^D} f(|c|).
%    \label{eq:coordinate}
%\end{equation}
by optimizing over the path assignments for item $v$ while fixing the assignments of all other items. Notice that the term $-\log N_v$ is irrelevant to $\pi_j(v)$ and hence can be dropped. The partial objective function related to item $v$ can be simplified as
%By further approximate $\log \sum_{j=1}^J s[v, \pi_j(v)]$ by $\sum_{j=1}^J \log s[v, \pi_j(v)]$ (this approximation seems unreasonable but leads a simpler algorithm which works well in practice), the parts of~\eqref{eq:coordinate} related to a particular item $v$ can be written as
\begin{align*}
    \arg\max_{\{\pi_{j}(v)\}_{j=1}^J} N_v \log \left( \sum_{j=1}^J s[v, \pi_j(v)] \right) - \alpha \sum_{j=1}^J f(|\pi_j(v)|).
\end{align*}
%so the optimal $\pi_j(v)$'s are simply the top $J$ values of $\log s[v, \pi_j(v)] - \alpha \left( f(|\pi_j(v)|+1) - f(|\pi_j(v)|) \right)$. 
Now we choose the path assignments $\{\pi_1(v), \dots, \pi_j(v)\}$ step by step. At step $i$, the incremental gain of the objective function by choosing $c = \pi_i(v)$ is given as follows:
\begin{align*}
    &&N_v \left( \log (\sum_{j=1}^{i-1} s[v, \pi_j(v)] + s[v, c]) - \log (\sum_{j=1}^{i-1} s[v, \pi_j(v)]) \right) \,\notag\\
    &&- \alpha \left( f(|c|+1) - f(|c|) \right).
\end{align*}
By keeping track of the partial sum $\sum_{j=1}^{i-1} s[v, \pi_j(v)]$ and the size of paths $|c|$, we greedily choose the path with the largest incremental gain. The details of the coordinate descent algorithm is given in Algorithm~\ref{alg:coodinate-descent}. In practice, three to five iterations are enough to ensure the algorithm converges. The time complexity grows linearly with vocabulary size $V$, multiplicity of paths $J$ as well as number of candidate paths $S$. 

\begin{figure*}[ht]
  \centering
\begin{minipage}{.72\linewidth}
\removelatexerror
\begin{algorithm}[H]
   {\bf Input}: Score functions $\log s[v, c]$. Number of iterations $T$.\\
   {\bf Initialization}: Set $|c| = 0$ for all paths $c$.\\
   \For{$t=1$ {\bfseries to} $T$} {
        \For{all items $v$}{
        Initialize the partial sum, ${\rm sum} \leftarrow 0$.\\
            \For{$j=1$ {\bfseries to} $J$}{
             \If{$t > 1$}{
                 $|\pi^{(t-1)}_j(v)| \leftarrow |\pi^{(t-1)}_j(v)| - 1$. (Reset path size for assignments from last iteration).
             }
             \For{all candidate paths $c$ of item $v$ such that $c \not\in \{\pi_l^{(t)}(v)\}_{l=1}^{j-1}$}{
             Compute incremental gain of objective function $$\Delta s[v, c] = N_v \left(\log ( s[v, c] + {\rm sum}) - \log({\rm sum}) \right) - \alpha \left( f(|c|+1) - f(|c|) \right).$$
             }
            $\pi_j^{(t)}(v) \leftarrow \arg\max_c \Delta s[v, c]$. \\
            ${\rm sum} \leftarrow  {\rm sum} + s[v, \pi_j^{(t)}(v)]$ (Update partial sum). \\
            $|\pi_j^{(t)}(v)| \leftarrow |\pi_j^{(t)}(v)| + 1$ (Update path size). \\
            }
        }
   }
 {\bfseries Output:} path assignments $\{\pi^{(T)}_j(v)\}_{j=1}^J$.
 \vspace{0.1in}
 \caption{Coordinate descent algorithm for penalized path assignment.}
   \label{alg:coodinate-descent}
\end{algorithm}
\end{minipage}
\end{figure*}

\section{Experiments on public datasets}
\label{sec:exp}
In this section, we study the performance of DR on two public recommendation datasets: MovieLens-20M~\citep{harper2015movielens} and Amazon books~\citep{he2016ups,mcauley2015image}. We compare the performance of DR with brute-force algorithm, as well as several other recommendation baselines including tree-based models TDM~\citep{zhu2018learning} and JTM~\citep{zhu2019joint}. At the end of this section, we investigate the role of important hyperparameters in DR. 

\subsection{Datasets and Metrics}

{\bf MovieLens-20M}. This dataset contains rating and free-text tagging activities from a movie recommendation service called MovieLens. We use the 20M subset which were created by the behaviors of 138,493 users between 1995 and 2015. Each user-movie interaction contains a used-id, a movie-id, a rating between 1.0 to 5.0, as well as a timestamp.

In order to make a fair comparison, we exactly follow the same data pre-processing procedure as TDM. We only keep records with rating higher or equal to 4.0, and only keep users with at least ten reviews. After pre-processing, the dataset contains 129,797 users, 20,709 movies and 9,939,873 interactions. Then we randomly sample 1,000 users and corresponding records to construct the validation set, another 1,000 users to construct the test set, and other users to construct the training set. For each user, the first half of the reviews according to the timestamp are used as historical behavior features and the latter half are used as ground truths to be predicted. 

{\bf Amazon books.} This dataset contains user reviews of books from Amazon, where each user-book interaction contains a user-id, an item-id, and the corresponding timestamp. Similar to MovieLens-20M, we follow the same pre-processing procedure as JTM. The dataset contains 294,739 users, 1,477,922 items and 8,654,619 interactions. Please note that Amazon books dataset has much more items but sparser interactions than MovieLens-20M. We randomly sample 5,000 users and corresponding records as the test set, another 5,000 users as the validation set and other users as the training set. The construction procedures of behavior features and ground truths are the same as in MovieLens-20M. 

{\bf Metrics.} We use precision, recall and F-measure as metrics to evaluate the performance for each algorithm. The metrics are computed for each user individually, and averaged without weight across users, following the same setting as both TDM and JTM. We compute the metrics by retrieving top $10$ and $200$ items for each user in MovieLens-20M and Amazon books respectively.

{\bf Model and training.} Since the dataset is split in a way such that the users in the training set, validation set and test set are disjoint, we drop the user-id and only use the behavior sequence as input for DR. The behavior sequence is truncated to length of 69 if it is longer than 69, and filled with a placeholder symbol if it is shorter than 69. A recurrent neural network with GRU is utilized to project the behavior sequence onto a fixed dimension embedding as the input of DR. We adopt the multi-task learning framework, and rerank the items in the recalled paths by a softmax reranker. We train the embeddings of DR and softmax jointly for the initial two epochs, freeze the embeddings of softmax and train the embeddings of DR for two more epochs. The reason is to prevent overfitting of the softmax model. In the inference stage, the number of items retrieved from beam search is not fixed due to the differences of path sizes, but the variance is not large. Empirically we control the beam size such that the number of items from beam search is 5 to 10 times the number of finally retrieved items.

\begin{table}[t]
    \caption{Comparison of precision@10, recall@10 and F-measure@10 for DR, brute-force retrieval and other recommendation algorithms on MovieLens-20M.}
    \centering
    \small
    \begin{tabular}{|c||c|c|c|}
        \hline
        Algorithm & Precision@10 & Recall@10 & F-measure@10 \\
        \hline 
        Item-CF & 8.25\% & 5.66\% & 5.29\% \\
        YouTube DNN & 11.87\% & 8.71\% & 7.96\% \\
        TDM (best) & 14.06\% & 10.55\% & 9.49\% \\
        \hline
        {DR} & {20.58 $\pm$ 0.47\%} & {10.89 $\pm$ 0.32\%} & {12.32 $\pm$ 0.36\%} \\
        Brute-force & 20.70 $\pm$ 0.16\% & 10.96 $\pm$ 0.32\% & 12.38 $\pm$ 0.32\% \\
        \hline 
    \end{tabular}
    % \hspace{1.5in}
    %\vspace{-0.25in}
    \label{tab:movielens_20m}
\end{table}

\begin{table}[t]
    \caption{Comparison of precision@200, recall@200 and F-measure@200 for DR, brute-force retrieval and other recommendation algorithms on Amazon Books.}
    \centering
    \small
    \begin{tabular}{|c||c|c|c|}
        \hline
        Algorithm & Precision@200 & Recall@200 & F-measure@200 \\
        \hline 
        Item-CF & 0.52\% & 8.18\% & 0.92\% \\
        YouTube DNN & 0.53\% & 8.26\% & 0.93\% \\
        TDM (best) & 0.56\% & 8.57\% & 0.98\% \\
        JTM & 0.79\% & 12.45\% & 1.38\% \\
        \hline
        {DR} & {0.95 $\pm$ 0.01\%} & {13.74 $\pm$ 0.14\%} & {1.63 $\pm$ 0.02\%} \\
        Brute-force & 0.95 $\pm$ 0.01\% & 13.75 $\pm$ 0.10\% & 1.63 $\pm$ 0.02\% \\
        \hline
    \end{tabular}
    %\hspace{1.5in}
    %\vspace{0.15in}
    \label{tab:amazon_books}
\end{table}

\subsection{Empirical Results}
We compare the performance of DR with the following algorithms: Item-CF~\citep{sarwar2001item}, YouTube product DNN~\citep{covington2016deep}, TDM and JTM. We directly use the numbers of Item-CF, Youtube DNN, TDM and JTM from TDM and JTM papers for fair comparison. Among the different variants of TDM presented, we pick the one with best performance. The result of JTM is only available for Amazon books.
We also compare DR with brute-force retrieval algorithm, which directly computes the inner-product of user embedding and all the item embeddings learnt in the softmax model and returns the top $K$ items. The brute-force algorithm is usually computationally prohibitive in practical large recommendation systems, but can be used as an upper bound for small dataset for inner-product based models.

\begin{table}[t]
\caption{Comparison of inference time for DR and brute-force retrieval on Amazon Books.}
\centering
\begin{tabular}{|c|c|}
    \hline
    {Deep Retrieval} & 0.266 ms per instance \\
    \hline
    Brute-force & 1.064 ms per instance\\
    \hline 
\end{tabular}
\vspace{-0.05in}
\label{tab:inference_time}
\end{table}

Table~\ref{tab:movielens_20m} shows the performance of DR compared to other algorithms and brute-force for MovieLens-20M. Table~\ref{tab:amazon_books} shows the results for Amazon books. For DR and brute-force, we independently train the same model for 5 times and compute the mean and standard deviation of each metric. Results are as follows.
\begin{itemize}
    \item DR performs better than other methods including tree-based retrieval algorithms such as TDM and JTM.
    \item The performance of DR is very close to or on par with the performance of brute-force method, which can be thought as an upper bound of the performance of vector-based methods such as Deep FM and HNSW. However, the inference speed of DR is 4 times faster than brute-force in Amazon books dataset (see Table~\ref{tab:inference_time}).
\end{itemize}

\subsection{Sensitivity of Hyperparameters}

\begin{figure*}[t]
    \centering
    \includegraphics[width=0.245\textwidth]{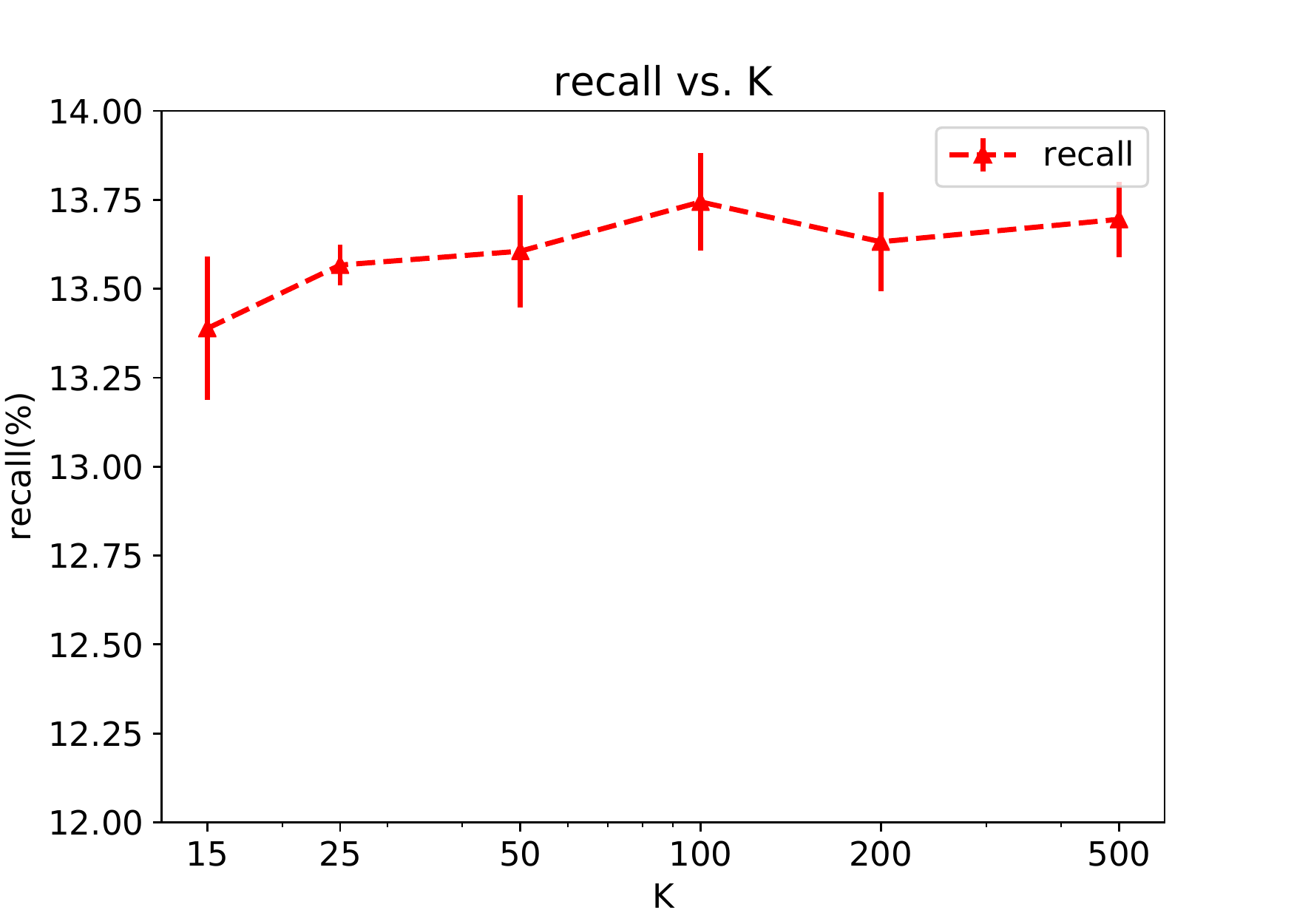}
    \includegraphics[width=0.245\textwidth]{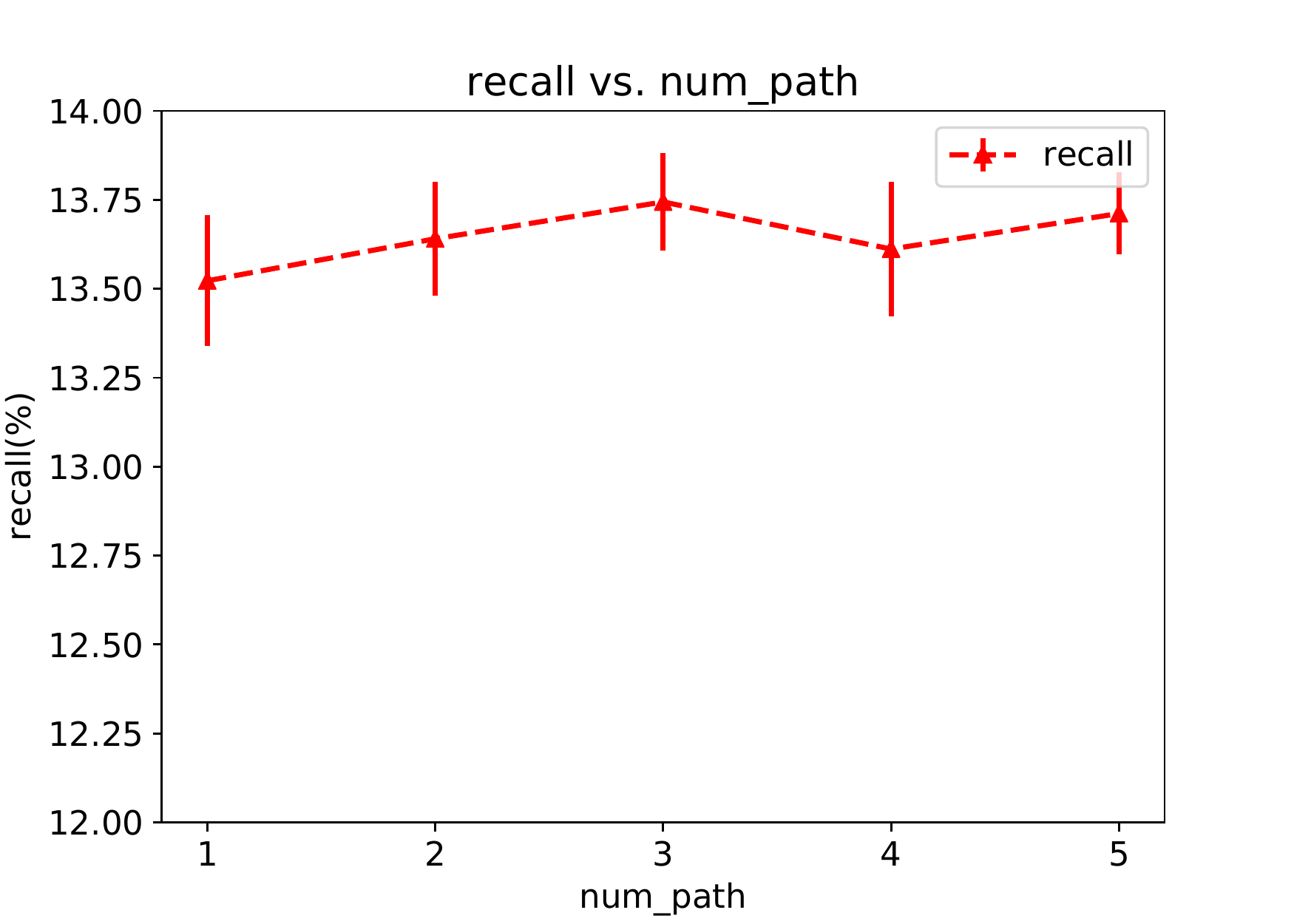}
    \includegraphics[width=0.245\textwidth]{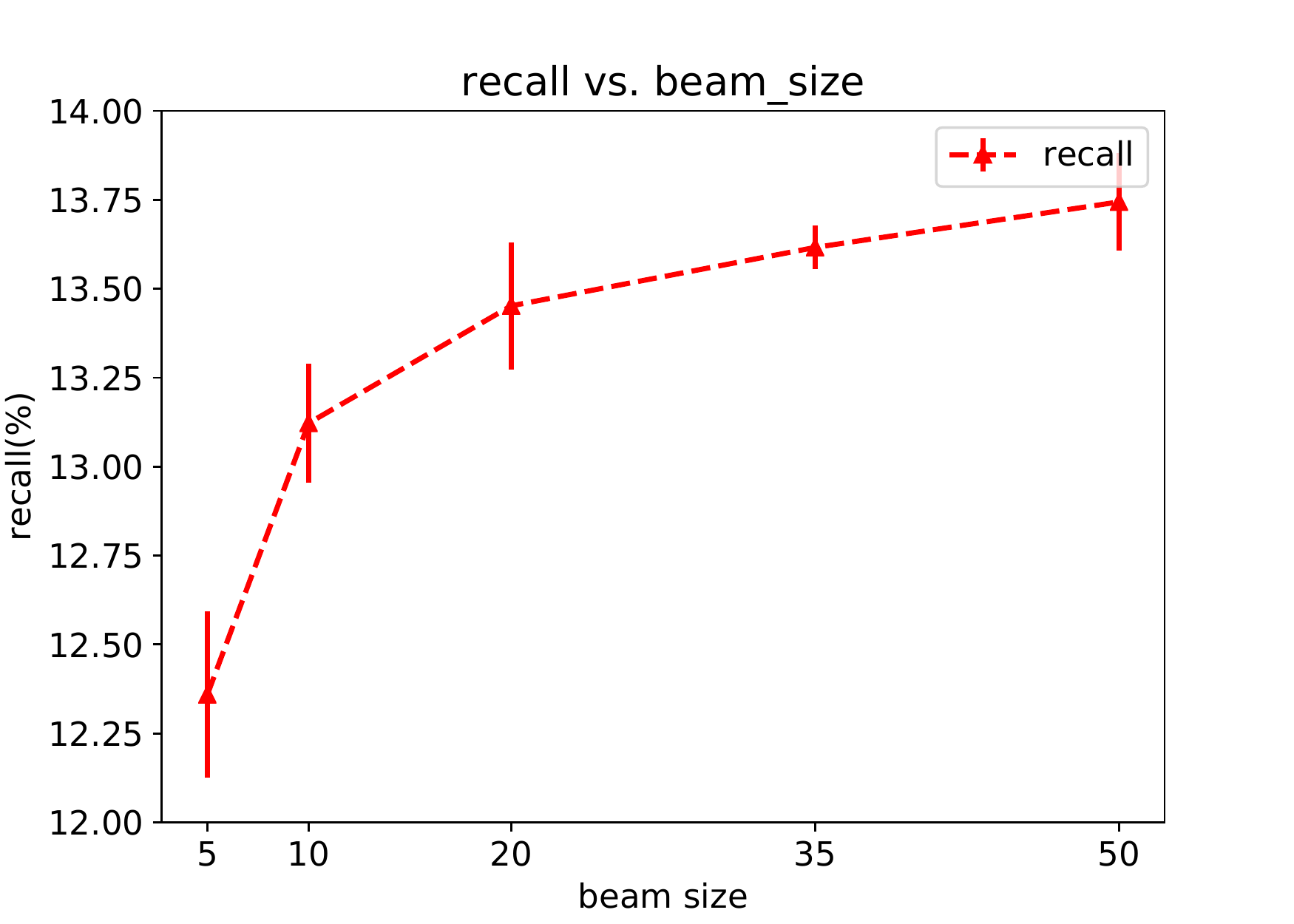}
    \includegraphics[width=0.245\textwidth]{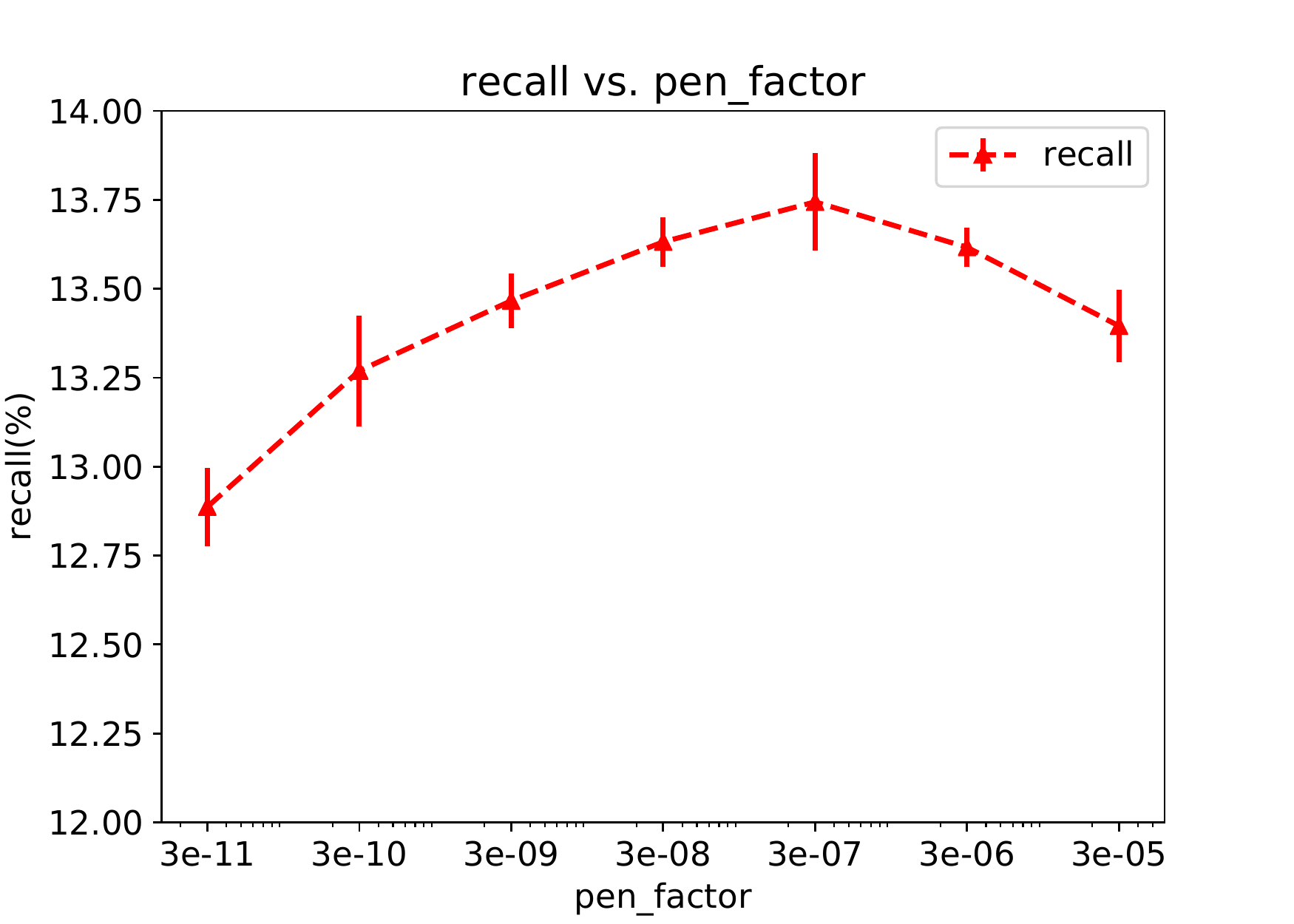}
    \caption{Relationship between recall@200 in Amazon Books experiment and model width $K$, number of paths $J$, beam size $B$ and penalty factor $\alpha$, respectively.}
    \label{fig:hyperparameters}
\end{figure*}

\begin{figure*}[t]
    \centering
    \includegraphics[width=0.245\textwidth]{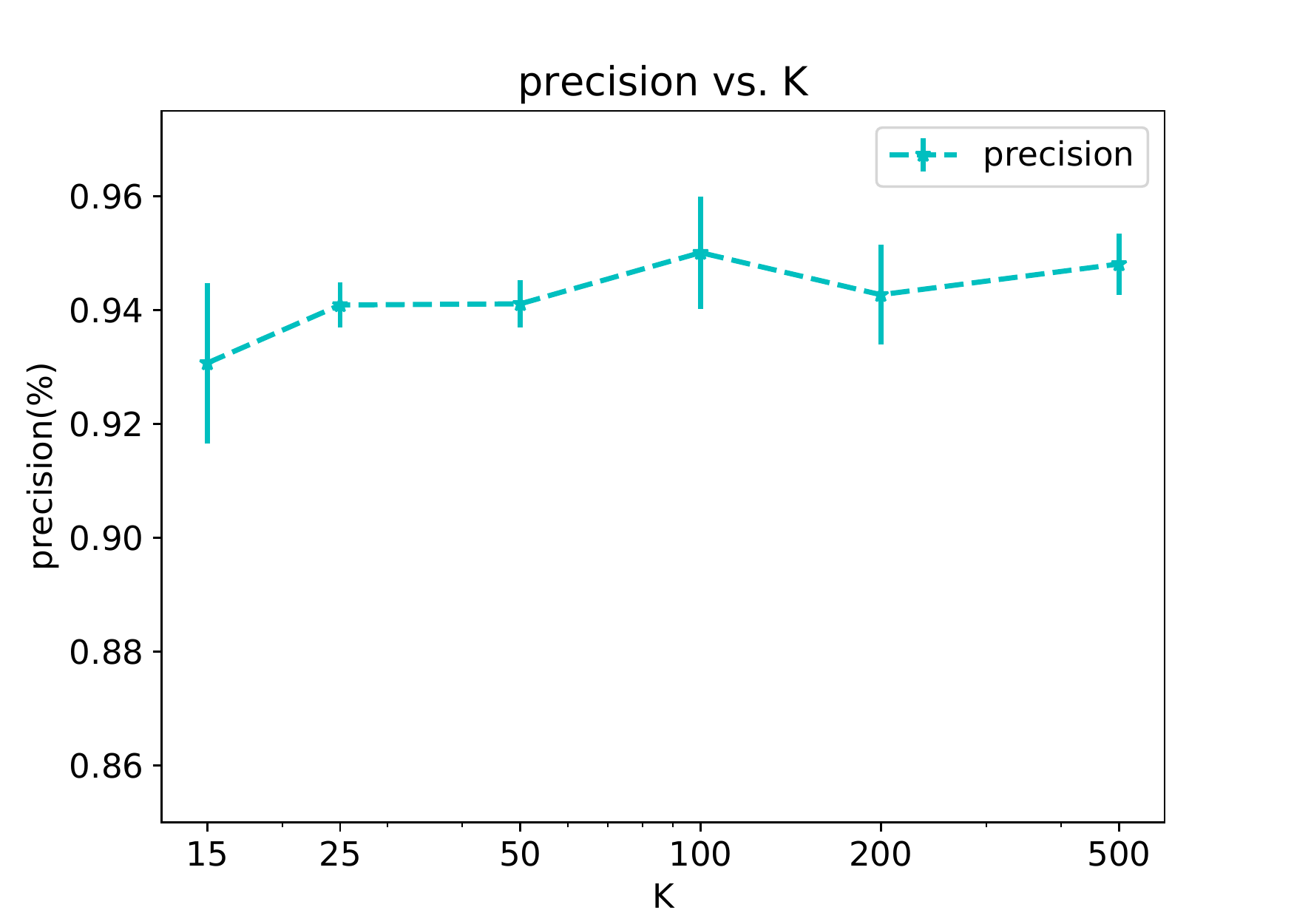}
    \includegraphics[width=0.245\textwidth]{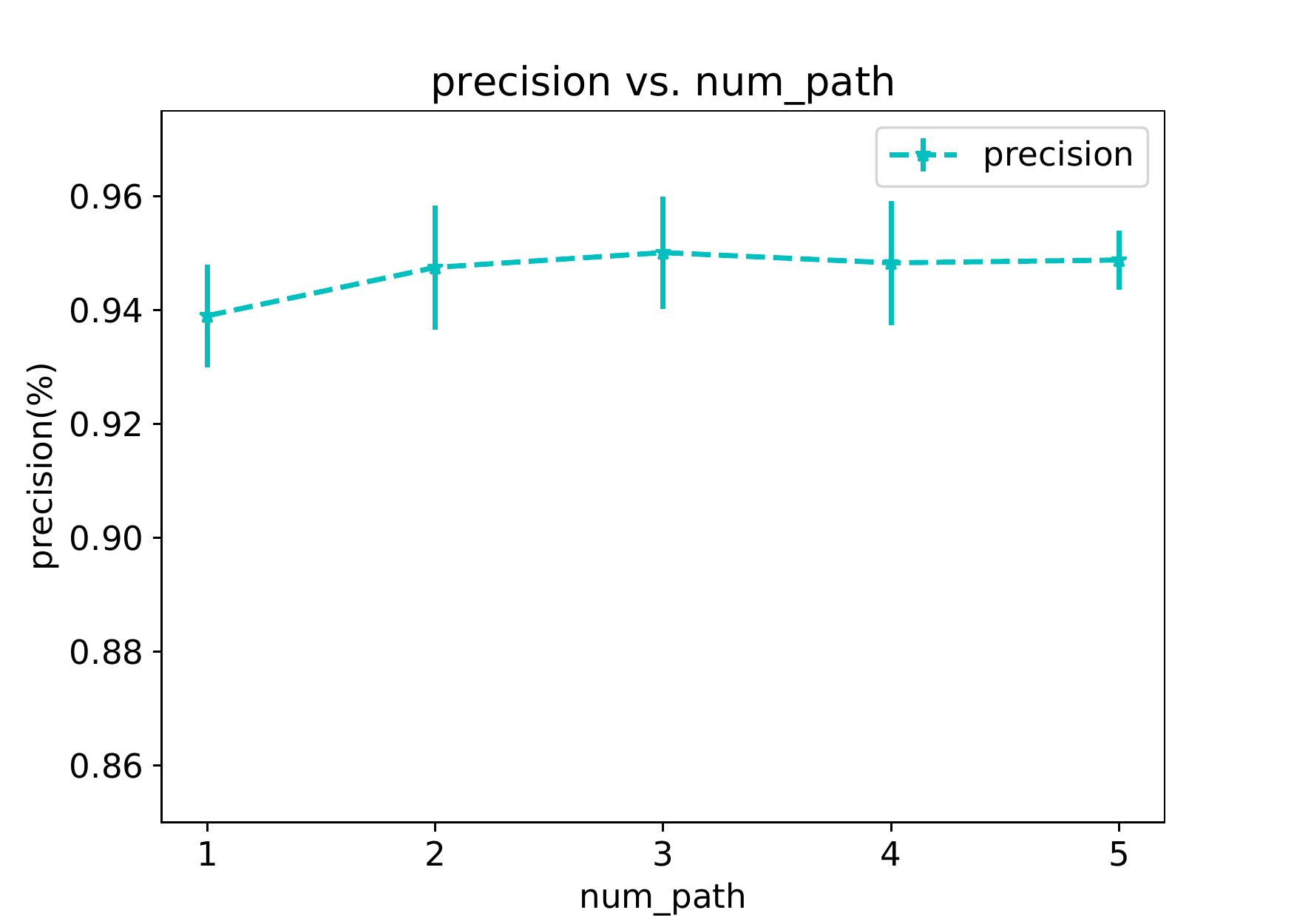}
    \includegraphics[width=0.245\textwidth]{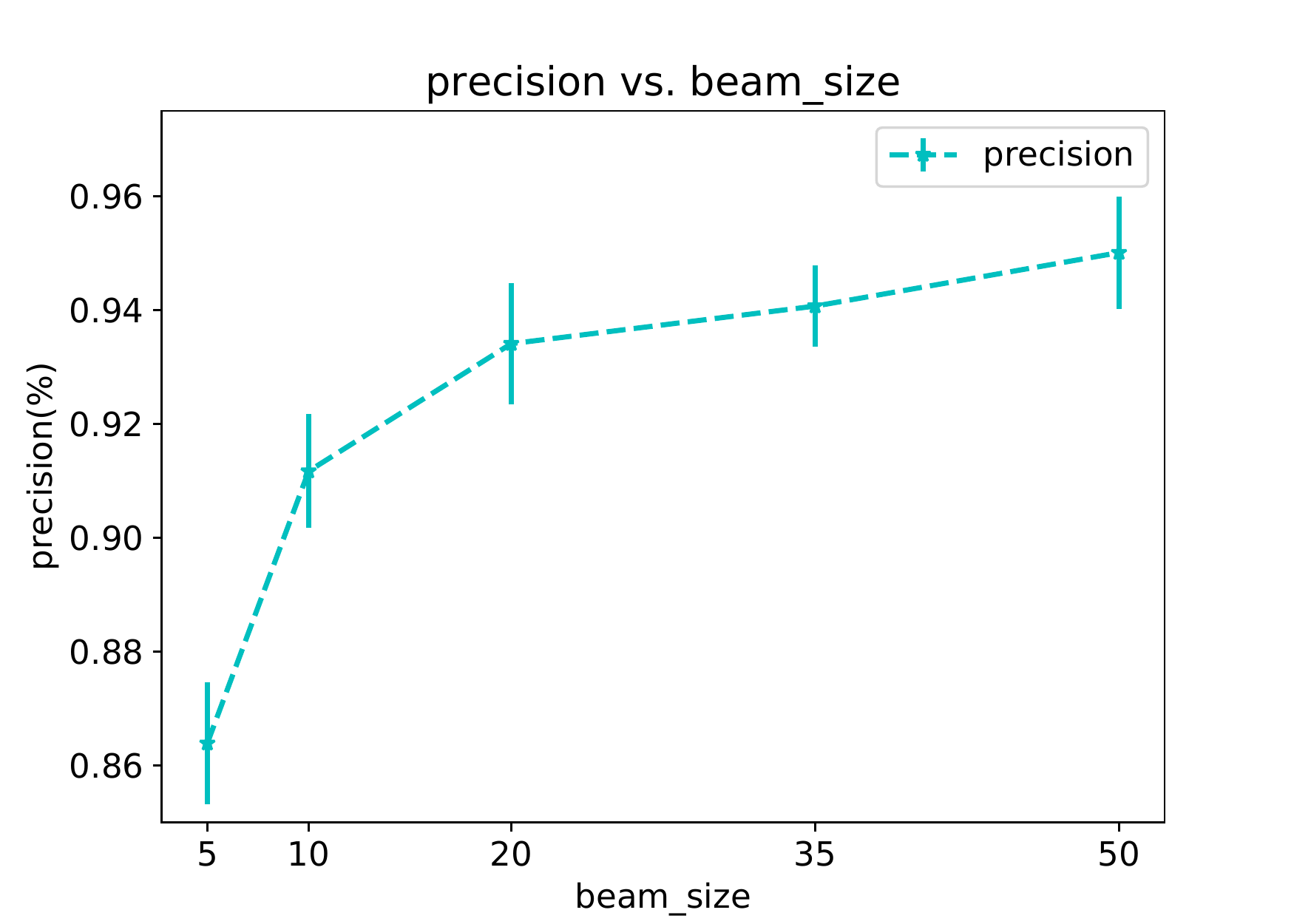}
    \includegraphics[width=0.245\textwidth]{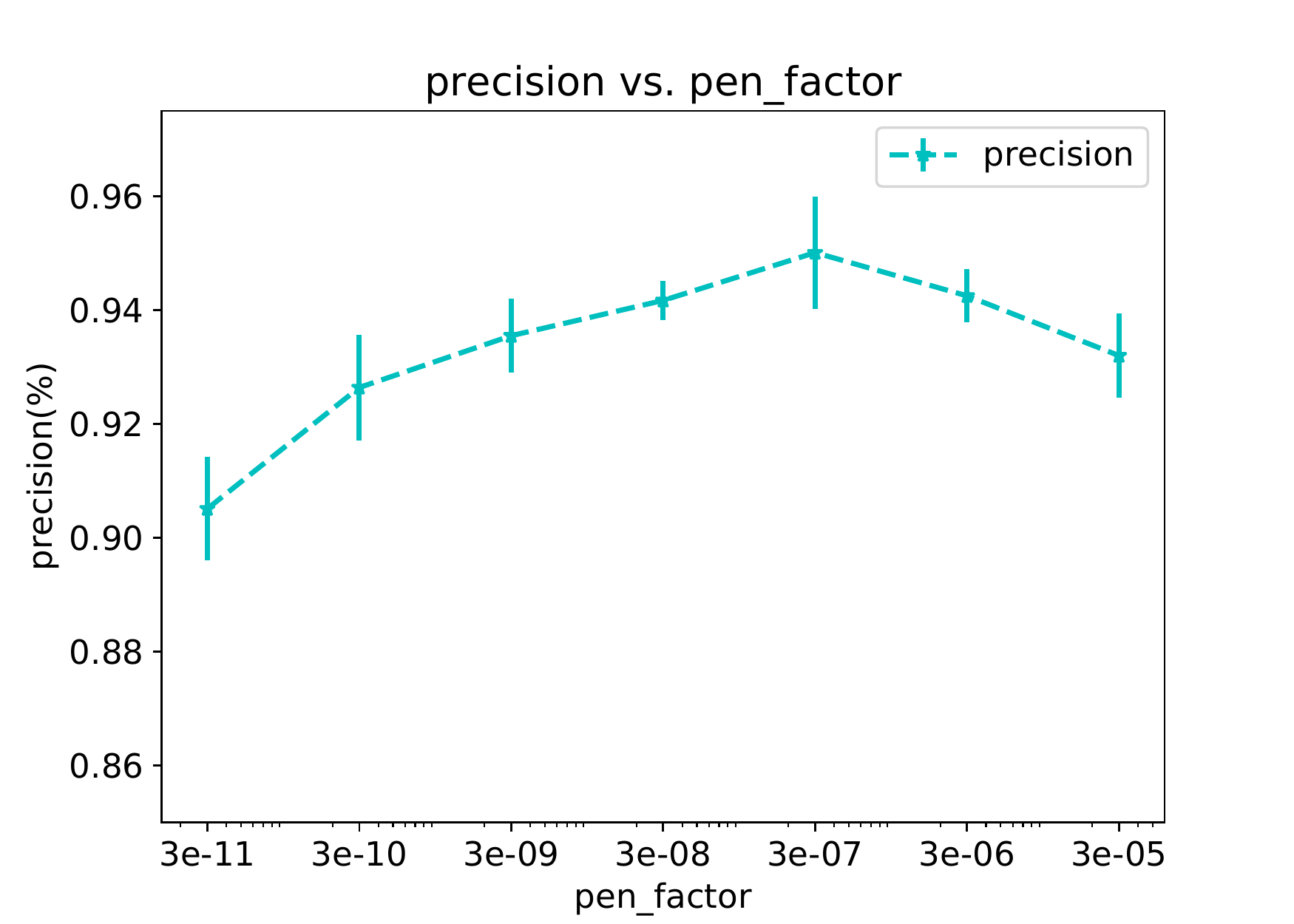}
    \caption{Relationship between precision@200 in Amazon Books experiment and model width $K$, number of paths $J$, beam size $B$ and penalty factor $\alpha$, respectively.}
    \label{fig:hyperparameters_2}
\end{figure*}

\begin{figure*}[t]
    \centering
    \includegraphics[width=0.245\textwidth]{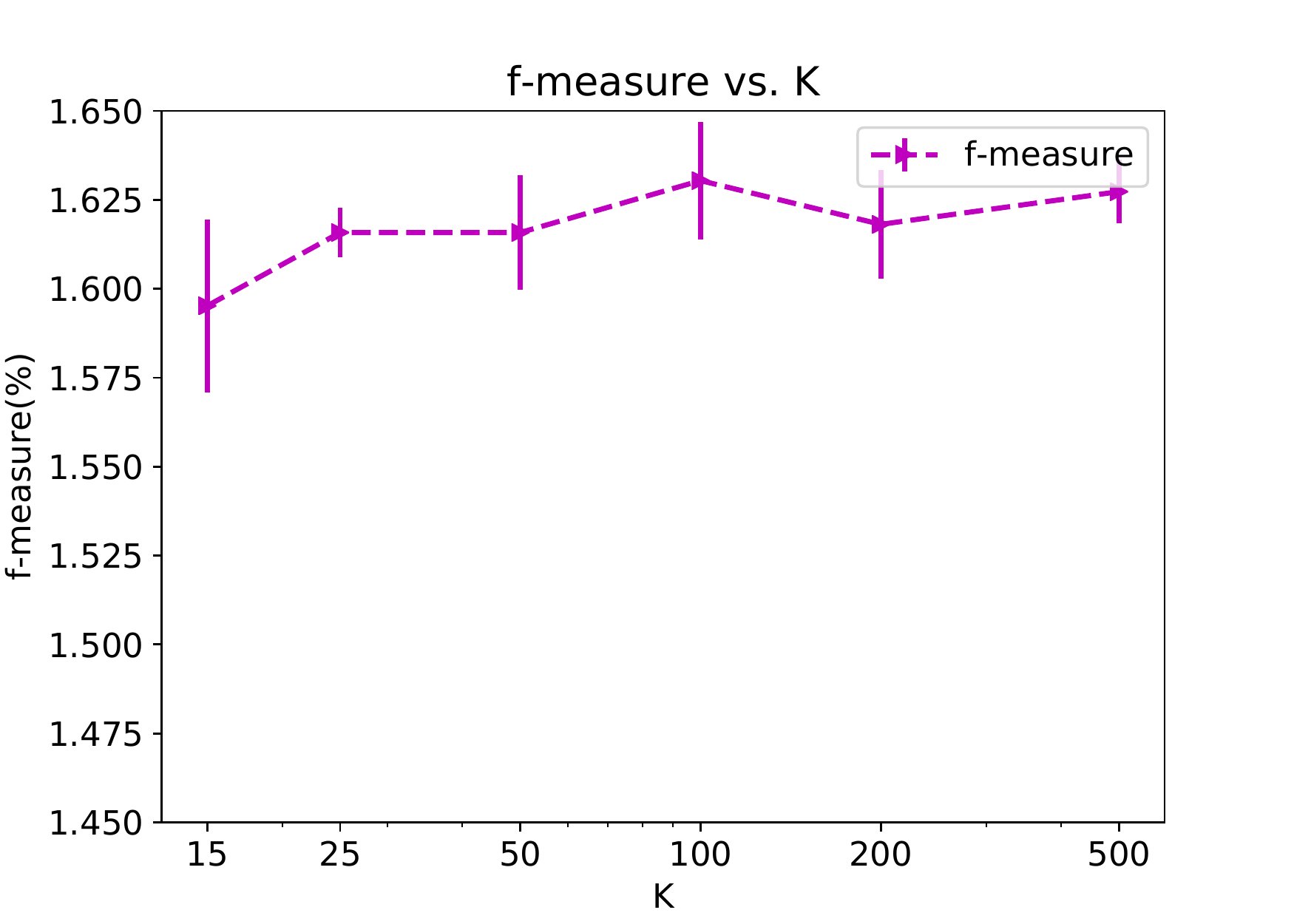}
    \includegraphics[width=0.245\textwidth]{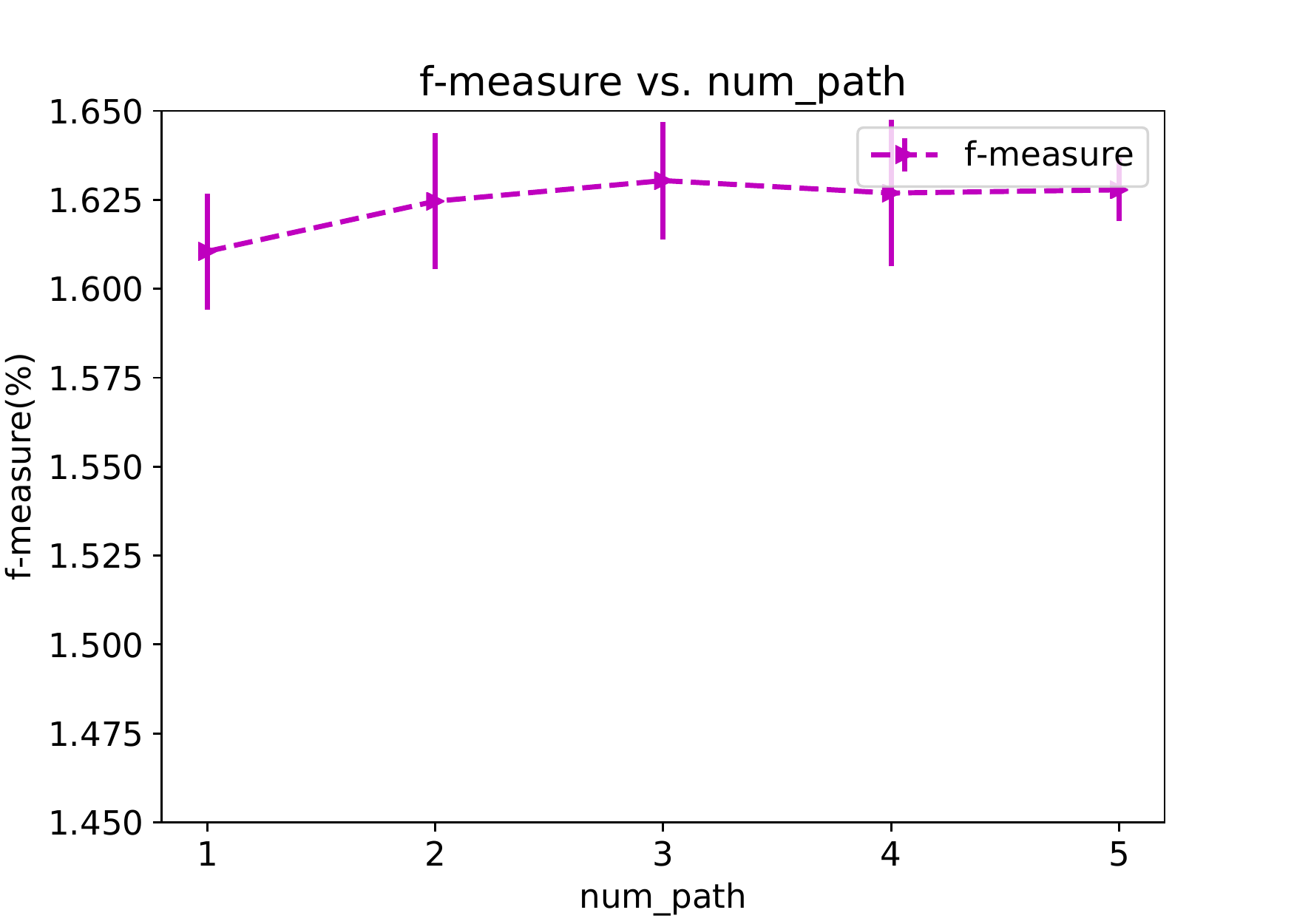}
    \includegraphics[width=0.245\textwidth]{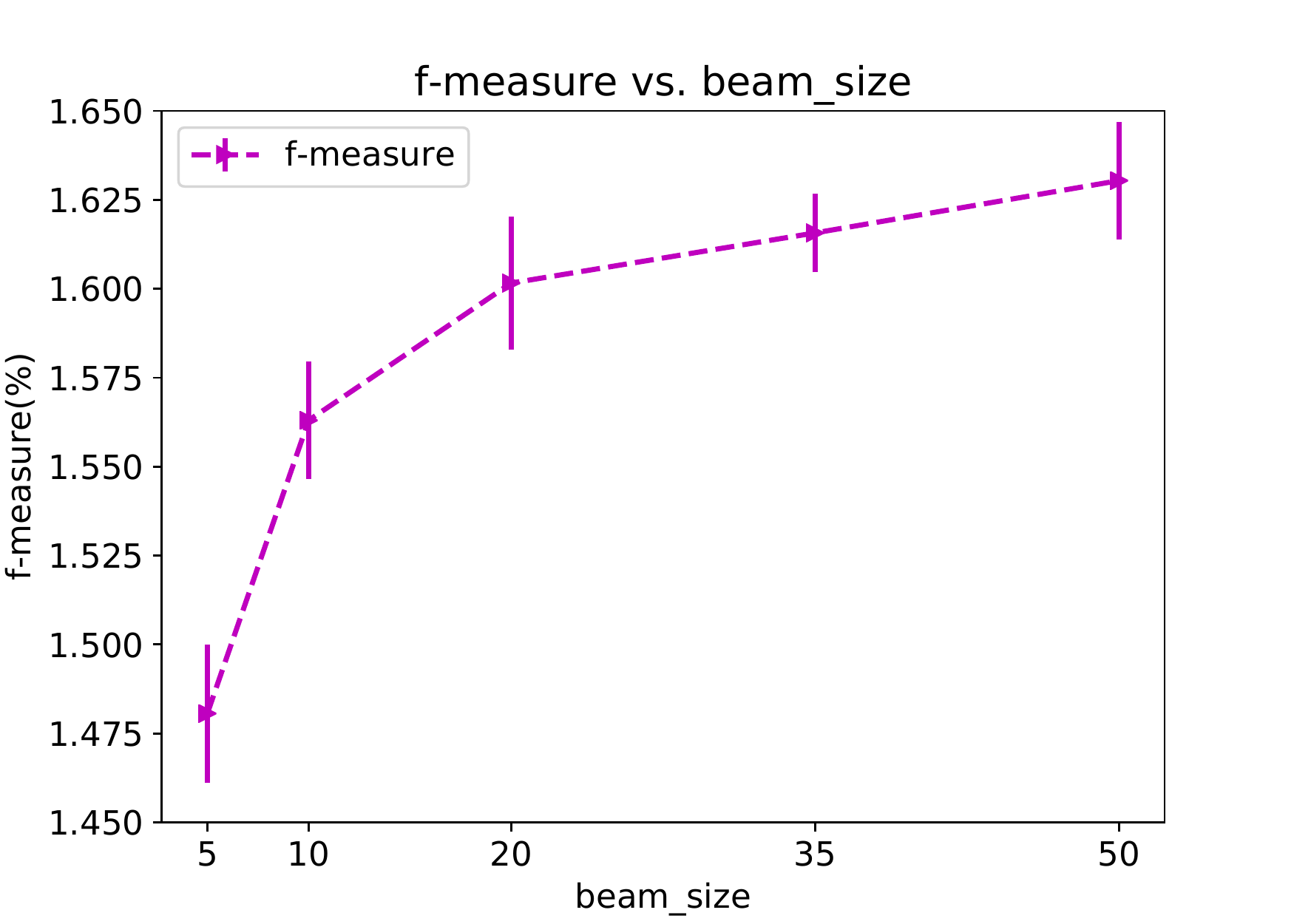}
    \includegraphics[width=0.245\textwidth]{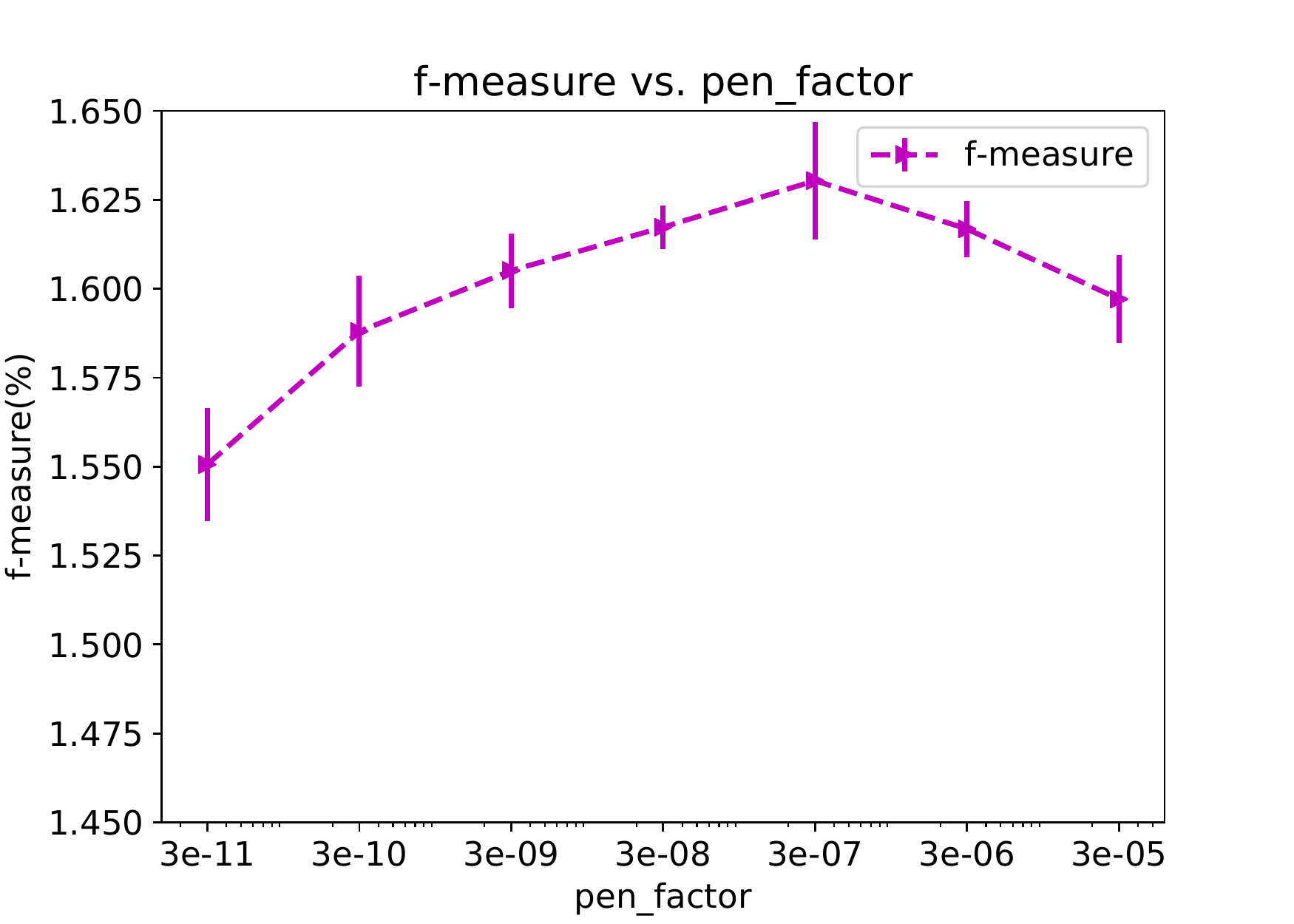}
    \caption{Relationship between F-measure@200 in Amazon Books experiment and model width $K$, number of paths $J$, beam size $B$ and penalty factor $\alpha$, respectively.}
    \label{fig:hyperparameters_3}
\end{figure*}

DR introduces some key hyperparameters which may affect the performance dramatically, including the width of the structure model $K$, depth of model $D$, number of multiple paths $J$, beam size $B$ and penalty factor $\alpha$. In the MovieLens-20M experiment, we choose $K=50$, $D=3$, $B=25$, $J=3$ and $\alpha = 3 \times 10^{-5}$. In the Amazon books experiment, we choose $K=100$, $D=3$, $B=50$, $J=3$ and $\alpha = 3 \times 10^{-7}$. Using the Amazon books dataset, we show the role of these hyperparameters and see how they may affect the performance. We present how the recall@200 change as these hyperparameters change in Fig.~\ref{fig:hyperparameters}. We keep the value of other hyperparameters unchanged when varying one hyperparameter. Precision@200 and F-measure@200 follow similar trends are shown in Fig.~\ref{fig:hyperparameters_2} and Fig.~\ref{fig:hyperparameters_3}.

\begin{table}[t]
\caption{Comparison of performance for different model depth $D$ on Amazon Books.}
    \centering
    \begin{tabular}{|c||c|c|c|}
        \hline
        $(K, D)$ & Precision @ 200 & Recall @ 200 & F-measure @ 200 \\
        \hline
        (100, 2) & 0.93\% & 13.34\% & 1.59\% \\
        (100, 3) & 0.95\% & 13.74\% & 1.63\% \\
        (100, 4) & 0.95\% & 13.67\% & 1.63\% \\
        (1000, 2) & 0.95\% & 13.68\% & 1.62\% \\ 
        \hline 
    \end{tabular}
    % \hspace{1.5in}
    %\vspace{0.15in}
    \label{D}
\end{table}

\begin{table}[t]
    \caption{Relationship between the size of the path with most items (top path size) and penalty factor $\alpha$.}
    \centering
    \small
    \begin{tabular}{|c||c|c|c|c|}
        \hline
         Penalty factor $\alpha$ & 3e-9 & 3e-8 & 3e-7 & 3e-6\\
         \hline
         Top path size & 1948 $\pm$ 30 & 956 $\pm$ 29 & 459 $\pm$ 13 & 242 $\pm$ 1 \\
         \hline
    \end{tabular}
    %\vspace{0.15in}
    \label{tab:top_path}
\end{table}

\begin{itemize}
    \item {\bf Width of model} $K$ controls the overall capacity of the model. If $K$ is small, the number of clusters is small for all the items; if $K$ is big, the time complexity of training and inference stages grow linearly with $K$. Moreover, large $K$ may increase the possibility of overfitting. An appropriate $K$ should be chosen depending on the size of the corpus.
    \item {\bf Depth of model} $D$. Using $D=1$ is obviously not a good idea since it fails to capture dependency among layers. In Table~\ref{D}, we investigate the result for $D=2, 3$ and $4$ for Amazon books dataset and conclude that (1) using $D=2$ with the same $K$ would hurt performance; (2) using $D=4$ with the same $K$ would not help. (3) Models with the same number of possible paths ($K=1000, D=2$ and $K=100, D=3$) have similar performance. However the number of parameters is of order $KD^2$, so a deeper model can achieve the same performance with fewer parameters. As a trade-off between model performance and memory usage, we choose $D=3$ in all experiments.
    \item {\bf Number of paths} $J$ enables the model to express multi-aspect information of candidate items. The performance is the worst when $J=1$, and keeps increasing as $J$ increases. Large $J$ may not affect the performance, but the time complexity for training grows linearly with $J$. In practice, choosing $J$ between 3 and 5 is recommended.
    \item {\bf Beam size} $B$ controls the number of candidate paths to be recalled. Larger $B$ leads a better performance as well as heavier computation in the inference stage. Notice that greedy search is a special case when $B=1$, whose performance is worse than beam search with larger $B$.
    \item {\bf Penalty factor} $\alpha$ controls the number of items in each path. The best performance is achieved when $\alpha$ falls in a certain range. From table~\ref{tab:top_path}, we conclude that smaller $\alpha$ leads to a larger path size hence heavier computation in the reranking stage. Beam size $B$ and penalty factor $\alpha$ should be appropriately chosen as a trade off between model performance and inference speed.
\end{itemize}

Overall, we can see that DR is fairly stable to hyperparameters since there is a wide range of hyperparameters which leads to near-optimal performances.

% \begin{itemize}
%     \item Amazon data experiment. Compare to JTM and other algorithms(Table 1). (In Appendix, Figures 3-5 showing how hyperparameters(K, beam size, penalty factor) affect the performance)
%     \item Toutiao data experiment. DR could have better recall than softmax for large beam size. (Table 2 or Figure 6) Softmax scales linearly on number of items, whereas DR sublinear.
% \end{itemize}

\section{Live experiments}
\label{sec:live-exp}
In this section, we show how DR works in real production systems. To the best of our knowledge, DR is among the first non-ANN algorithms successfully deployed at the scale of hundreds of millions of items for industrial recommendation systems. 

While the results on public datasets in the last section shed light upon the basic behavior of the DR models, it is more important to understand if it could improve user experiences for real recommendation systems in industrial environments. These systems are much more complicated and difficult to improve due to sheer volume of the data (user-generated contents, UGC) and their dynamic nature---there are new items uploaded almost every second. 
%In our experiences, it is very much possible that an algorithm works for some public datasets (at the scale of millions), but does not improve for the real systems with orders of magnitude larger data.

\begin{figure}[t]
    \centering
    %\vspace{-0.2in}
    \includegraphics[width=0.5\textwidth]{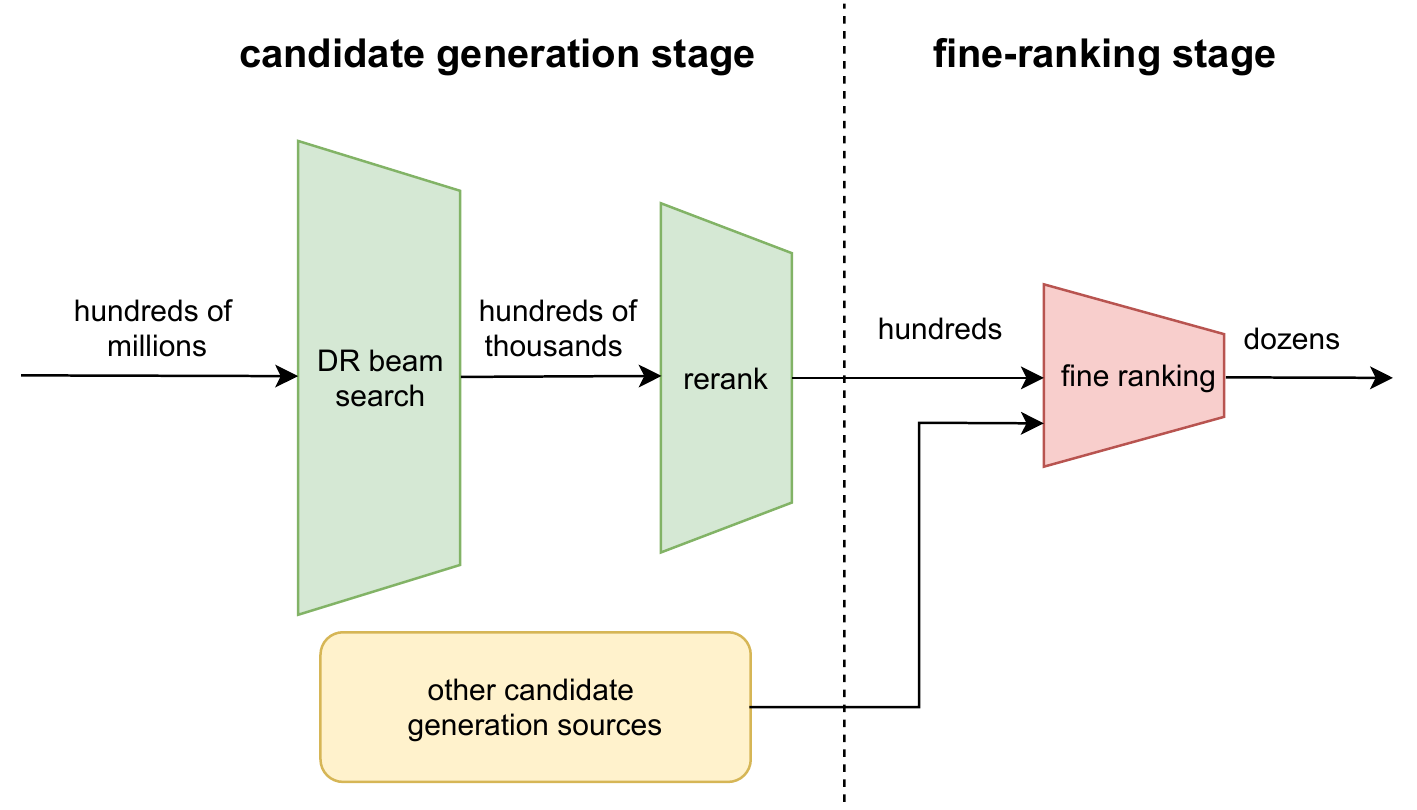}
    %\vspace{-0.3in}
    \caption{An illustrative diagram to show how to use DR in an industrial recommendation system.}
    \label{fig:rec-sys}
    \vspace{-0.1in}
\end{figure}

Since the number of items in our system is usually in the order of hundreds of millions, making the full-scale fine ranking a difficult task. Thus, an industrial-scale recommendation system usually consists of several stages as shown in the YouTube recommendation paper~\cite{covington2016deep}. 
At minimal, the system has two stages: candidate generation and fine-ranking. Here fine-ranking stage usually uses a more complex model with additional features that are computationally intractable in the candidate generation stage. Thus fine ranking usually handles only hundreds of items for each user request.  DR works as the one of candidate generation components. Fig.~\ref{fig:rec-sys} shows the illustrative diagram of this process. In practice, multiple candidate generation sources are used at the same time to produce a diverse set of candidates for the fine-ranking stage. 

Our experimental platform is one of the largest in the industry with hundreds of millions of users and items. In our system, items are short videos. For this particular DR experiment, the labels are generated to indicate whether a user has finished watching the video or not. The baseline candidate generation method used Field-aware Factorization Machine (FFM)~\citep{juan2016field} to learn the user and item embeddings and followed by HNSW~\citep{malkov2018efficient} for approximate nearest neighbor search. This baseline was extensively tuned to maximize the user experiences on the platform. Slightly different from the use of softmax for reranking on public datasets for DR, the re-ranking model we use here is a logistic regression model. Previous attempts of using softmax in the production did not result in meaningful improvements. The reranking model architecture is the same for both HNSW and DR. 

We report the performance of DR in AB test in Table~\ref{tab:live_exp}. DR enjoys a significant gain in key metrics, including video finish rate, app view time and second day retention, over the baseline model. All improvements are statistically significant. This is likely because the end-to-end training of DR aligns the learning of the user embeddings and the retrievable structure under the same objective function directly from user-item interactions. So the clustering of DR contains more user-behavior information of candidate items. We also found that DR is more friendly to less popular videos or less popular creators, and the AB test shows that much more of such videos are recommended to the end users. This is beneficial to the platform's creator ecosystem. We believe the reason is as follows. Within each path in the DR structure, items are indistinguishable and this allows the less popular items to be retrieved as long as they share some similar behaviors with popular ones.
Last but not least, DR is naturally suitable for streaming training, and the time to build the structure for retrieval is much less than HNSW since there is no computation of any user or item embeddings involved in DR's M-step. It takes about 10 minutes to process the total items with a multi-thread CPU implementation. In industrial recommendation systems like ours, these improvements are substantial. This experiment shows the advantage of DR in large-scale recommendation systems.
\begin{table}[t]
    \caption{Live experiment results on different engagement metrics.}
    \centering
    \begin{tabular}{|c|c|}
        \hline
        Metric & Relative improvement of DR \\ \hline
         Video finish rate & +3.0\%  \\ \hline
         App view time & +0.87\% \\ \hline
         Second day retention & +0.036\% \\\hline
    \end{tabular}
    \label{tab:live_exp}
\end{table}

\section{Conclusion and discussion}
\label{sec:discussion}
In this paper, we have proposed Deep Retrieval, an end-to-end learnable structure model for large-scale recommendation systems. DR uses an EM-style algorithm to learn the model parameters and paths of items jointly. Experiments have shown that DR performs well compared with brute-force baselines in two public recommendation datasets as well as live production environments with hundreds of millions of users and items.	
There are several future research directions based on the current model design. Firstly, the structure model defines the probability distribution on paths only based on user side information. A useful idea would be how to incorporate item side information more directly into the DR model. 
%and use the similarity of probability distribution on paths based on user embeddings and item embeddings as objective to jointly train the two models. 
Secondly, in the structure model, we only make use of positive interactions such as click, convert or like between user and item. Negative interactions such as non-click, dislike or unfollow should also be considered in future work to improve the model performance. Finally, we currently use a simple dot-product type of models as a reranker and we plan to use a more complex model in the future.

%\begin{acks}
%To Robert, for the bagels and explaining CMYK and color spaces.
%\end{acks}

\balance
\bibliographystyle{ACM-Reference-Format}
\bibliography{kdd}


% \appendix

% \section*{Appendix}
% \label{sec:append}
% \input{appendix}

\end{document}